\documentclass[preprint,12pt,authoryear]{elsarticle}

 \usepackage{graphicx}

\usepackage{amssymb}

 \usepackage{lineno}

 \journal{Journal of Atmospheric and Solar Terrestrial Physics}

\begin{document}

\begin{frontmatter}



\title{Hydromagnetic Waves in a Compressed Dipole Field via Field-Aligned Klein-Gordon Equations}


\author[aa]{Jinlei Zheng}
\author[bb,cc]{Qiang Hu\corref{cor1}}\ead{qh0001@uah.edu}
\author[cc,dd]{J. F. McKenzie}
\author[cc]{G. M. Webb}

\cortext[cor1]{Corresponding author. Tel.: +1-256-961-7920; fax:
+1-256-961-7730.}
\address[aa]{Department of Physics, University of
Alabama, Huntsville, AL, USA.} \address[bb]{Department of Space
Science, University of Alabama, Huntsville, AL, USA.}
\address[cc]{Center for Space Plasma and Aeronomic Research
(CSPAR), University of Alabama, Huntsville, AL, USA.}
\address[dd]{Department of Mathematics and Statistics, Durban
University of Technology, Steve Biko Campus, Durban, South
Africa.}

\begin{abstract}
Hydromagnetic waves, especially those of frequencies in the range
of a few milli-Hz to a few Hz observed in the Earth's
magnetosphere, are categorized as Ultra Low Frequency (ULF) waves
or pulsations. They have been extensively studied due to their
importance in the interaction with radiation belt particles and in
probing the structures of the magnetosphere. We developed an
approach in examining the toroidal standing Aflv\'{e}n waves in a
background magnetic field by recasting the wave equation into a
Klein-Gordon (KG) form along individual field lines. The
eigenvalue solutions to the system are characteristic of a
propagation type when the corresponding eigen-frequency is greater
than a cut-off frequency and an evanescent type otherwise. We
apply the approach to a compressed dipole magnetic field model of
the inner magnetosphere, and obtain the spatial profiles of
relevant parameters and the spatial wave forms of harmonic
oscillations. We further extend the approach to poloidal mode
standing Alfv\'{e}n waves along field lines. In particular, we
present a quantitative comparison with a recent spacecraft
observation of a poloidal standing Alfv\'{e}n wave in the Earth's
magnetosphere. Our analysis based on KG equation yields consistent
results which agree with the spacecraft measurements of the wave
period and the amplitude ratio between the magnetic field and
electric field perturbations. We also present computational
results of eigenvalue solutions to the compressional poloidal mode
waves in the compressed dipole magnetic field geometry.
\end{abstract}


\end{frontmatter}


\section{Introduction}
\label{intro}
 Hydromagnetic waves are common phenomena in space plasmas. The associated magnetic and electric field
 perturbations
 are observed both on ground and from space in the Earth's
 magnetosphere. Such waves or magnetic pulsations of frequencies
 less than $\sim 1$ Hz are typically categorized as Ultra Low
 Frequency (ULF) waves \citep{2006GMSF,2006GMSK}. They can be further divided into
 sub-categories, such as Pc1-5, Pi1-3 and Pg, with frequencies ranging from a
 few Hz down to a few milli-Hz (mHz) \citep{2006GMSF,2006GMSV}.
 Generally they  exhibit regular and monochromatic  magnetic and electric field wave forms. Such waves can be
 identified as Alfv\'{e}n waves propagating in the Earth's
 magnetosphere, e.g., the recent spacecraft observation by the Van Allen Probe
(Radiation Belt Storm Probes)  of
 \citet{2013GeoRLD}. Based on the direction or the polarization of
 the magnetic (or electric) field perturbation in the linearized assumption, they can be
 further characterized as  toroidal  and/or poloidal mode
 waves. In the toroidal mode, the magnetic field perturbation is
 in the azimuthal direction, i.e., along the east-west longitudinal
 direction (the accompanying electric field perturbation has a radial component) in Earth's dipole magnetic
 field. On the other hand in the poloidal mode, the magnetic field perturbation has
 a radial component, lying in the meridional plane, while the
 electric field perturbation is azimuthal.

 The basic theory for ULF waves traces back to \citet{1965Tamao} and has
 been well developed. These waves are interpreted   as standing
 Alfv\'{e}n (transverse) or fast-mode hydromagnetic waves in cold
 plasmas immersed in the Earth's magnetic field
 \citep{1969JGRC,1981JGRS,1983SSRvS}. Their characteristics are
 closely governed by the geometry of the background magnetic field
 and the associated plasma  density distribution. More general and sophisticated numerical
 simulations were also developed in recent years to take into
 account more realistic background field topology, multiple
 physical effects, and non-idealized boundary conditions \citep{2007AnGeoK,2006GMSL,2010JGRAC}.
 The study of ULF waves has important implications for
 wave-particle interaction and diagnostics of magnetospheric
 structures. In particular, it has been established the critical role that ULF waves play in
 the energization and transport of radiation belt particles based
 on both theoretical and observational studies \citep{2006GMSE,2003JGRAE,1999GeoRLE,2002JGRAT,2005JGRAU}.

An alternative approach to describing the toroidal (transverse)
Alfv\'{e}n standing waves in an axi-symmetric background magnetic
field has been given by \citet{2010AnGeoM} where the wave
equations were cast along an individual field line and transformed
into a Klein-Gordon (KG) form. This approach was further
formalized and applied to the Earth's dipole magnetic field. We
later showed in great detail the formulation and procedures of the
approach for a given background field topology and density
distribution in \citet{2012JGRAW}. The eigen-frequencies obtained
from the eigen-mode solutions to the KG equations correspond well
to the ULF waves frequencies in the Pc3-5 range \citep{2012JGRAW}.
The same approach was also successfully applied to coronal loop
oscillations in low corona under different background field and
plasma conditions \citep{2012AAH}. In the present work, we first
apply the approach to a more realistic Earth's background field as
represented by a compressed dipole model of \citet{2007AnGeoK}. We
derive the eigen-frequencies and eigen-functions of the wave forms
for this particular geometry corresponding to the toroidal mode
and compare with the results from other similar studies.

Furthermore, motivated by a recent direct observation of poloidal
 standing Alfv\'{e}n waves in Earth's magnetosphere by
\citet{2013GeoRLD} \citep[see
also,][]{2013JGRAT,2013JGRAL,2011GeoRLL}, we extend our
investigation to examine the poloidal mode waves  as well. In the
case of a transverse poloidal mode, the wave equation can also be
cast into a KG form along a field line. We numerically solve the
wave equation for electric field perturbation. The corresponding
magnetic field perturbation can be obained in a similar manner to
the approach based on KG equations for the toroidal mode. In
\citet{2013GeoRLD}, a case of a fundamental mode standing poloidal
wave was identified from the newly launched Van Allen Probe
(Radiation Belt Storm Probes) spacecraft measurements. They
obtained  wave period of the azimuthal electric field and the
associated radial magnetic field oscillations, the relative ratio
of wave amplitudes, and the relative phase shift at the spacecraft
location in the inner magnetosphere. Their analysis provided clear
evidence for the existence of poloidal mode waves and their
interaction with particles.

The article is organized as follows. Section~\ref{gov} provides a
brief summary of the toroidal-mode wave equations and their
transformation into the KG form. The general approach of solving
the resulting eigenvalue problem is described and applied to a
compressed dipole magnetic field model of Earth's magnetosphere.
The eigen-frequencies and the corresponding wave-form solutions
are presented. Section~\ref{pol} extends the analysis to the
decoupled eigenvalue solutions of the poloidal mode waves for a
given geometry, and presents a comparison with the solution from
other approaches and with observations, especially for the
transverse Alfv\'{e}n waves. We finally summarize our results in
the last section.

\section{Klein-Gordon Equations for the Toroidal Mode}\label{gov}
We first consider toroidal wave perturbations ($b_\phi, u_\phi$)
in the magnetic field and fluid velocity in a background
axi-symmetric ({that is azimuthal wave number $m=0$}) magnetic
field $\vec{B}_0=(B_r, B_\theta, 0)$, in  spherical coordinates
$(r,\theta,\phi)$. The perturbation electric field $\vec{E}$ is
given by
\begin{equation}
\vec{E}=-\vec{u} \times\vec{B}=-u_\phi B_r \hat{\theta}+u_\phi
B_\theta \hat{r}=E_n\hat{n},\label{eqE}
\end{equation}
normal to the background magnetic field line. The $\phi$
(toroidal) components of Faraday's law and the momentum equation
yield the following wave equations for the perturbations, when
evaluated along individual field lines that can be specified by a
functional form $r(\theta)$ between the radial distance $r$ and
the co-latitude $\theta$ \citep{2010AnGeoM,2012JGRAW}:
 \begin{eqnarray}
 \frac{\partial^2 b_\phi}{\partial
 t^2}&=&\frac{V^2}{r^2}\left\{\frac{d^2
 b_\phi}{d\theta^2}-\frac{1}{L_b}\frac{d
 b_\phi}{d\theta}+\frac{b_\phi}{M_b}\right\}\label{eqwvb}\\
\frac{\partial^2 u_\phi}{\partial
 t^2}&=&\frac{V^2}{r^2}\left\{\frac{d^2
 u_\phi}{d\theta^2}-\frac{1}{L_u}\frac{d
 u_\phi}{d\theta}+\frac{u_\phi}{M_u}\right\},\label{eqwvu}
 \end{eqnarray}
where $V=B_\theta/\sqrt{\mu_0\rho}$ with a given background plasma
density $\rho$ and all coefficients are functions of $\theta$
only. The total derivative is given
\begin{equation}
\frac{d}{d\theta}=\frac{\partial}{\partial\theta}+\frac{rB_r}{B_\theta}\frac{\partial}{\partial
r}.\label{ddth}\end{equation} These waves equations are to be
solved along individual field lines by being transformed into
(linear) Klein-Gordon equations of ordinary differential equation
type. The solutions are obtained for a given background magnetic
field topology and the associated density distribution, and
harmonic time dependence, subject to specific boundary conditions.
The detailed derivation, formulations and procedures are given in
\citet{2012JGRAW}, including a case study of a standard dipole
field. We restrict our presentation mostly to a brief description
of the general case (see below).

\subsection{General Case}\label{sec:general}
The  perturbations of physical quantities as given by
equations~(\ref{eqwvb}) and (\ref{eqwvu}) have the general form
\begin{equation}
\frac{\partial^2\Psi}{\partial
t^2}=\frac{V^2}{r^2}\left[\frac{d^2\Psi}{d\theta^2}
-\frac{1}{\mathcal{L}}\frac{d\Psi}{d\theta}+\frac{1}{\mathcal{M}}\Psi\right]\label{eqKG}
\end{equation}
which can be transformed into the Klein-Gordon form through the
substitution
\begin{equation}
\Psi=\psi \exp\int\frac{d\theta}{2\mathcal{L}},
\quad \mathrm{where}\quad
\frac{1}{\mathcal{L}}=-\frac{d}{d\theta}\ln(F(\theta)).\label{eqL}
\end{equation}
This yields
\begin{equation}
\frac{\partial^2\psi}{\partial
t^2}+\omega_c^2\psi=\frac{V^2}{r^2}\frac{d^2\psi}{d\theta^2}\label{eqwvphi}
\end{equation}
in which a cut-off frequency $\omega_c$ is manifest and given by
\begin{equation}
\omega_c^2=\frac{V^2}{r^2}\left[\frac{1}{2\mathcal{L}^2}(1+\mathcal{L}')-\frac{1}{\mathcal{M}}\right].\label{eqwc2}
\end{equation}
The amplitude factor in Eq.~(\ref{eqL}) becomes
\begin{equation}
\exp\int\frac{d\theta}{2\mathcal{L}}=\frac{1}{F^{1/2}(\theta)}=f(\theta).\label{eqF}
\end{equation}
This factor arises from the adiabatic-geometric growth or decay
corresponding to conservation of wave energy flux through a flux
tube as given by Poynting's theorem \citep{2010AnGeoM}. For the
velocity perturbation, in particular, the relevant factor is
simply $F(\theta)=B_\theta/r$ \citep{2012JGRAW}. That the quantity
$\omega_c$, given by (\ref{eqwc2}), in Eq.~(\ref{eqwvphi}) is
indeed a cut-off frequency  is readily seen by taking a harmonic
time variation $\propto\exp(i\omega t)$ for then
Eq.~(\ref{eqwvphi}) becomes
\begin{equation}
\frac{d^2\psi}{r^2
d\theta^2}=-\frac{(\omega^2-\omega_c^2)}{V^2}\psi\equiv -k^2\psi
\label{eqphik2}
\end{equation}
An equation of this form possesses propagating-type solutions,
provided $\omega>\omega_c$ (or $\omega^2>\omega_c^2$) and
evanescent solutions for $\omega<\omega_c$. {If a slowly varying
background is assumed, JWKB solutions yield good approximations to
the propagating and evanescent behavior.} The imposition of
boundary conditions (e.g., at the end points of one field line)
yield an eigenvalue problem for $k$ (and hence $\omega$).

The procedures for solving the toroidal wave equations were given
by \citet{2012JGRAW} and \citet{2012AAH}. We adopt the usual
boundary condition $E_n\equiv 0$ (i.e., $u_\phi=0$), when solving
the eigenvalue problem for $u_\phi$ satisfying the KG
Eq.~(\ref{eqphik2}). Physically this corresponds to the situation
of the field-line footpoints rooted in the Earth's ionosphere of
infinite conductivity. The toroidal velocity perturbations are
then  obtained by solving the KG equation subject to the boundary
condition and the transformation of the growth/decay factor. A set
of the solutions of different wave forms is obtained for a
discrete set of eigenvalues $\omega$ which usually correspond to a
set of harmonic oscillations with increasing frequency and number
of nodes \citep{2012JGRAW}. Then the accompanying toroidal
magnetic field perturbation is calculated by
\begin{equation} \frac{\partial b_\phi}{\partial
t}=\frac{B_\theta}{r}\frac{du_\phi}{d\theta}+\frac{u_\phi
B_\theta}{r}\frac{1}{l_b},\label{eq:bphi}
\end{equation} where the function
$\frac{1}{l_b}=-\frac{d}{d\theta}\ln(r\sin\theta)$ is known once
the background magnetic field topology is given. Depending on the
specific eigen-mode solution being sought, a constant
eigen-frequency $\omega$ and the corresponding eigen-function
solutions are obtained for both $u_\phi$ and $b_\phi$.

\begin{figure}[htb]
\begin{center}
\includegraphics[width=8.3cm]{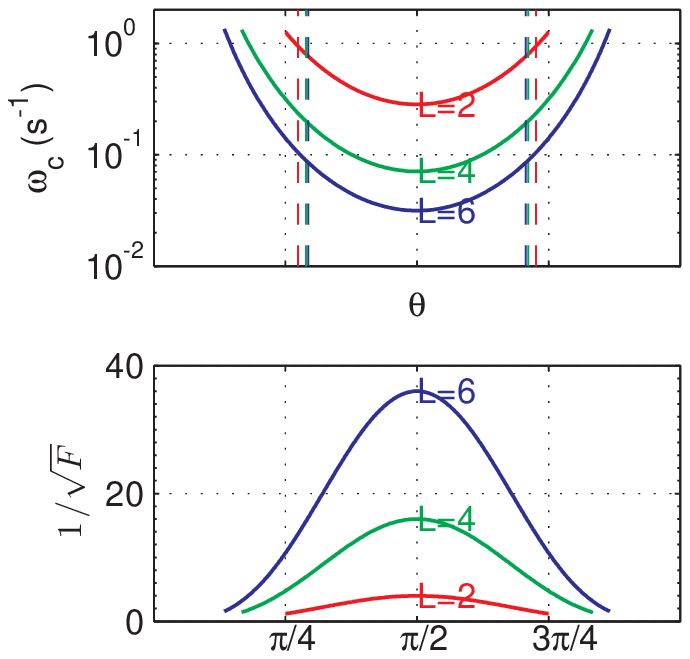}
\end{center}
\caption{The parameter $\omega_c$  (with $B_0=0.31$ Gauss,
$a=6.4\times 10^8$ cm,  and $\rho_e=7$ amu $ \mathrm{cm}^{-3} $),
and the adiabatic growth/decay factor as a function of $\theta$
for various $L$ values  of a dipole field. The vertical lines mark
the location (colatitude) where $\omega_c=\omega_0$, the
eigen-frequency of the corresponding fundamental
mode.}\label{fig00}
\end{figure}

As examples, the cases of a standard dipole field with a typical
power-law density distribution have been examined for ULF waves in
Earth's magnetosphere \citep{2012JGRAW} and coronal loop
oscillations in Sun's corona \citep{2012AAH} by the above
approach. Fig.~\ref{fig00} shows the variation of the cut-off
frequency $\omega_c$ and the amplitude factor $f(\theta)$ for an
axi-symmetric Earth's dipole field, particularly for $L=2,4,6$
(here the value $L$ as in ``$L$-shell" represents the radial
distance of one particular field line crossing the equator). We
use a density model by \citet{2007AnGeoK} throughout the present
study: $\rho=\rho_e\left(\frac{r}{5}\right)^4$ with $\rho_e=7$ amu
cm$^{-3}$ and the radial distance $r$ is measured in Earth radius.
The general profiles of $\omega_c$ and $f$ are similar to those
presented in \citet{2012JGRAW}, but their magnitudes are sensitive
to the different background density distributions assumed, as are
the eigen-frequencies obtained. Table~\ref{tbl:dpl} lists the
eigen-frequency of the fundamental mode $\omega_0$, the
corresponding period $T_0$, and locations $\theta_0$ along each
individual field line where $\omega_0=\omega_c(\theta_0)$ for the
dipole field. Given the profiles of $\omega_c(\theta)$ in
Fig.~\ref{fig00}, we find that for $\theta_0< \theta<
\pi-\theta_0$ where $\omega>\omega_c$, the solution of the
propagation type exists, while beyond that interval where
$\omega<\omega_c$, decaying type solution exists, as reflected in
the resulting wave forms from the corresponding eigen-function
solutions \citep[see][]{2012JGRAW}. The same set of results will
be obtained for the case of a compressed dipole field in the
following subsection.

\subsection{{A Compressed Dipole Field}}
A compressed dipole field is given in the spherical coordinate
(which is intrinsically a 3D field, but remains planar at each
$\phi$) \citep{2007AnGeoK}:
\begin{eqnarray}
B_r&=&\left(\frac{2B_0}{r^3}-b_1(1+b_2\cos\phi)\right)\cos\theta, \label{eqBr}\\
B_\theta&=&\left(\frac{B_0}{r^3}+b_1(1+b_2\cos\phi)\right)\sin\theta, \label{eqBth} \\
B_\phi&=&0. \label{eqBph}
\end{eqnarray}
So our approach can only be approximately applied to the
noon-midnight meridional planes corresponding to $\phi=0$ and
$\phi=\pi$ respectively, on which $\partial/\partial \phi=0$.

\begin{figure}[htb]
\begin{center}
\includegraphics[width=12cm]{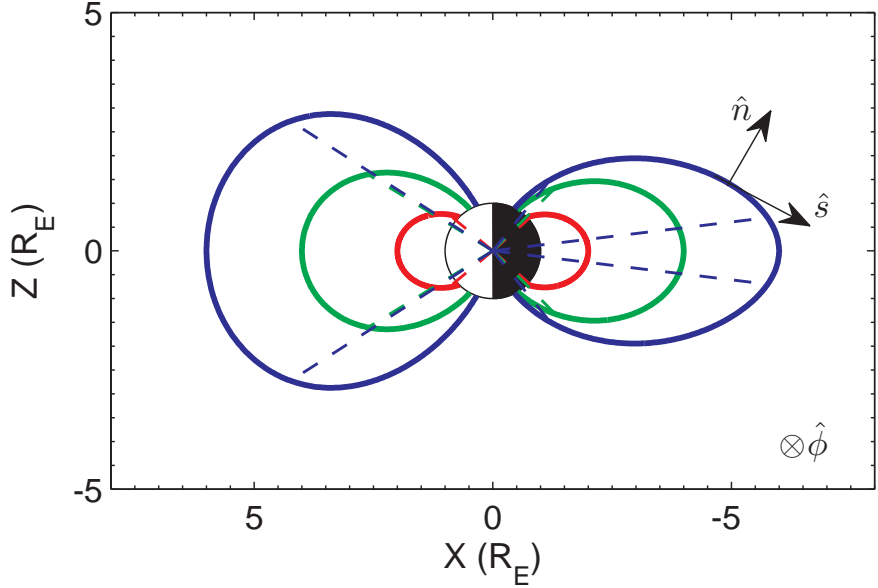}
\end{center}
\caption{The magnetic field lines of $L=2,4,6$ on the
noon-midnight meridional plane for the compressed dipole with
$B_0=0.31$ Gauss, $b_1=10$ nT, and $b_2=8$. The dashed lines mark
the locations where the eigen-frequencies of the fundamental mode
intersect the cut-off frequencies, i.e.,  $\omega_0=\omega_c$ as
shown in Fig.~\ref{fig1} and given in Tables~\ref{tbl:cdpl0} and
\ref{tbl:cdpl1}. The field-line aligned orthogonal coordinate $(n,
s, \phi)$ is also shown.}\label{fig0}
\end{figure}

Fig.~\ref{fig0} shows the selected field lines for $L=2,4$, and 6,
respectively, in both the noon ($\phi=0$) and midnight
($\phi=\pi$) meridional planes of the Earth as illustrated. The
asymmetry between the two sides is clearly seen due to the
compression of the solar wind on the noon side ($\mathrm{X}>0$).
We carry out the analysis of toroidal mode waves for each
individual field line via the approach of KG equations outlined in
Section~\ref{sec:general}. The corresponding formulations of
various coefficients, cut-off frequency and growth/decay amplitude
factor for the compressed dipole geometry are given in the
Appendix.

\begin{figure}[htb]
\begin{center}
\includegraphics[width=.45\textwidth]{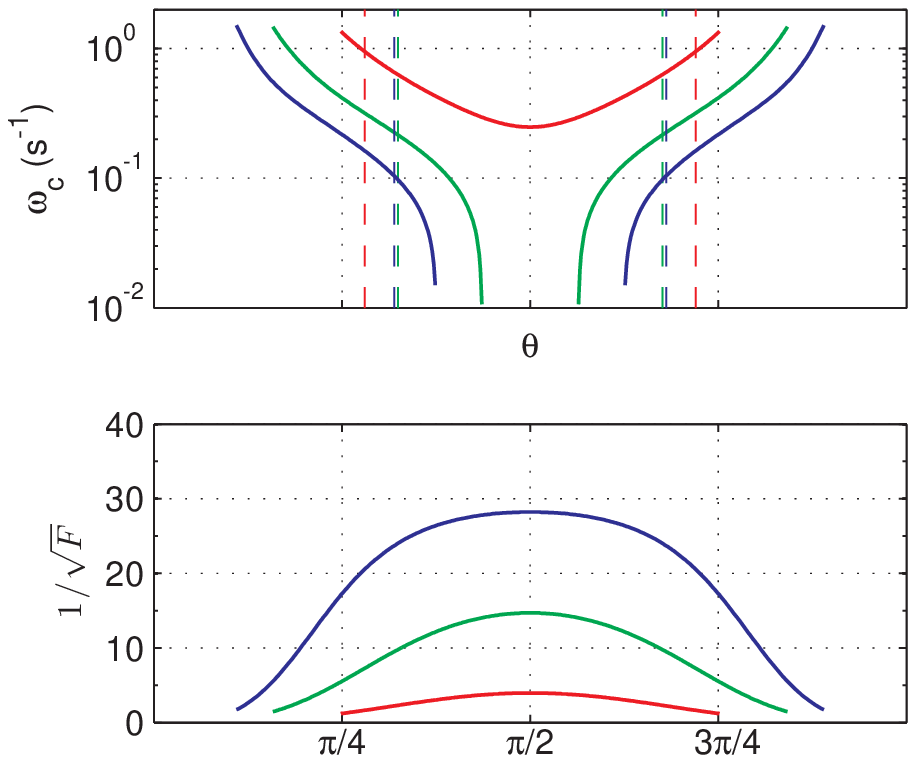}
\includegraphics[width=.45\textwidth]{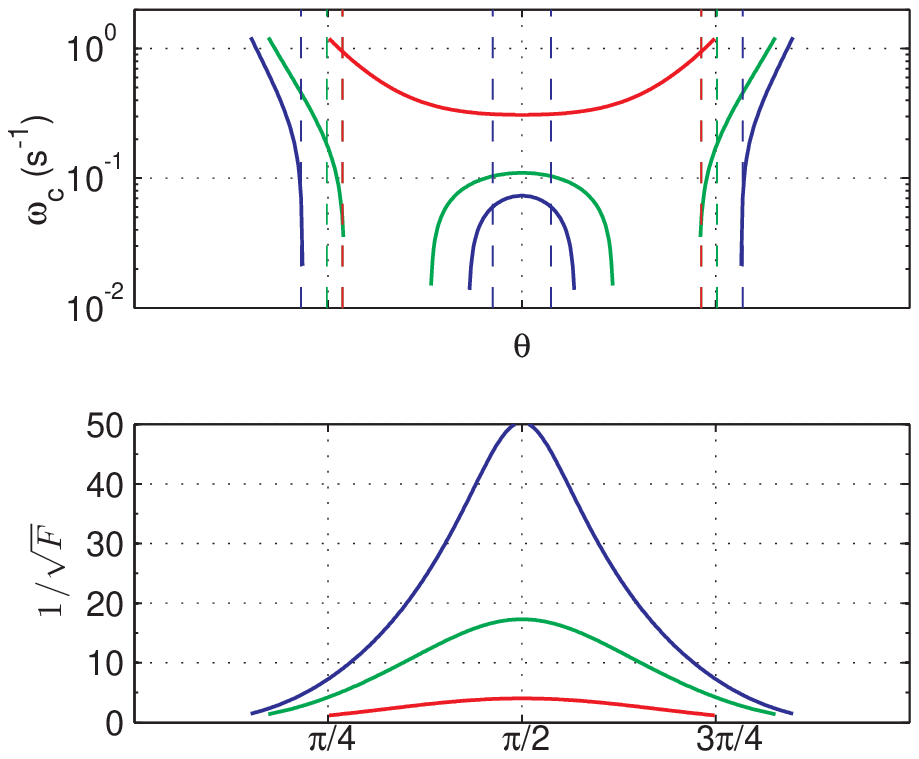}
\\\hspace{.07\textwidth} (a) \hspace{.41\textwidth} (b)
\end{center}
\caption{The parameter $\omega_c$  (with $B_0=0.31$ Gauss,
$a=6.4\times 10^8$ cm,  and $\rho_e=7$ amu cm$^{-3}$), and the
adiabatic growth/decay factor as a function of $\theta$ for
various $L$ values   for the compressed dipole field with (a)
$\phi=0$ and (b) $\phi=\pi$, respectively. Format is the same as
Fig.~\ref{fig00}. The broken part of some curves corresponds to
$\omega_c^2<0$.}\label{fig1}
\end{figure}

First of all, the profiles of cut-off frequency $\omega_c$ and the
amplitude factors are calculated and illustrated in
Fig.~\ref{fig1}, together with the locations where the
eigen-frequencies of the fundamental mode intersect the cut-off
frequencies. The corresponding parameters of the eigen-frequency
$\omega_0$, the period $T_0$, the co-latitude $\theta_0$, and the
radial distance $r_0$ where $\omega_c=\omega_0$ are given in
Tables~\ref{tbl:cdpl0} and \ref{tbl:cdpl1} for the noon and
midnight side, respectively. The profiles of $\omega_c$ and $f$
show significant differences among the cases of noon, midnight
meridional plane of the compressed dipole, and that of a standard
dipole, especially for greater $L$ values. For example, for the
case $L=6$ at midnight side, the amplitude factor peaks at a
greater value $\sim 50$ at the equator, while the eigen-frequency
$\omega_0$ intersects the cut-off frequency at two locations in
$\theta<\pi/2$, one near the north pole and the other near the
equator. Therefore there are two separate regions of propagating
solution to the KG equation where $\omega^2>\omega_c^2$ and one
additional region of evanescent solution surrounding the equator
as marked by the pairs of dashed blue lines along congruent points
in co-latitude. However for higher-order harmonics, the
eigen-frequency increases with the increasing number of nodes such
that it becomes greater than the cut-off frequency throughout the
whole range of low latitudes enclosing the equator.

Overall, the values of parameters for the fundamental mode are
comparable among the cases presented in
Tables~\ref{tbl:dpl}-\ref{tbl:cdpl1}, although significant
deviations also exist especially for the case of $L=6$. The
periods range between a few to tens of seconds and a little over
one-hundred seconds, with increasing $L$ values, which correspond
well to the frequency range of Pc1-5 ULF waves in Earth's
magnetosphere. For the compressed dipole cases, the periods also
agree well with those reported by \citet{2007AnGeoK}, where the
periods rose from a few seconds at $L=2$, to tens of seconds at
$L=4$, and to $\sim 100$ seconds at $L=6$. The heights (radial
distances) of the locations where $\omega_c=\omega_0$ increases
with $L$ values, reaching much greater values in the compressed
dipole case than that in the standard dipole. The corresponding
co-latitudes, on the other hand, remain close to each other,
except for the one near equator for $L=6$ in the midnight side of
the compressed dipole case.

\begin{figure}[htb]
\begin{center}
\includegraphics[width=.45\textwidth]{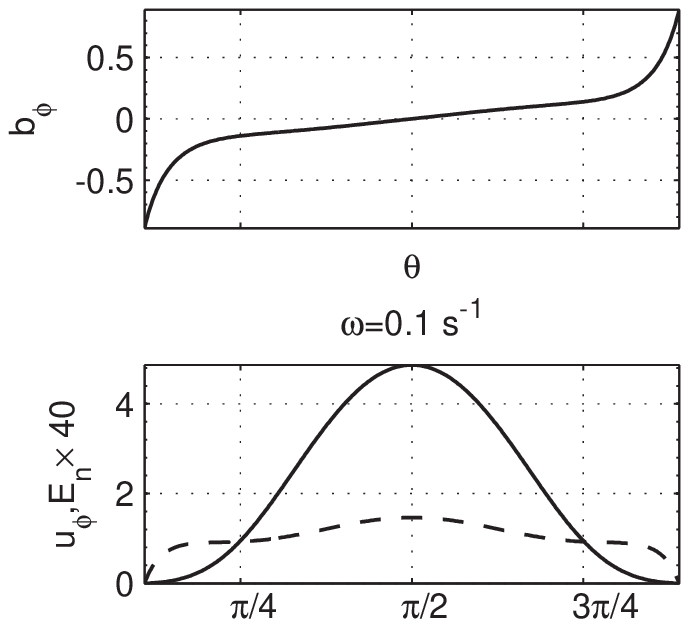}
\includegraphics[width=.45\textwidth]{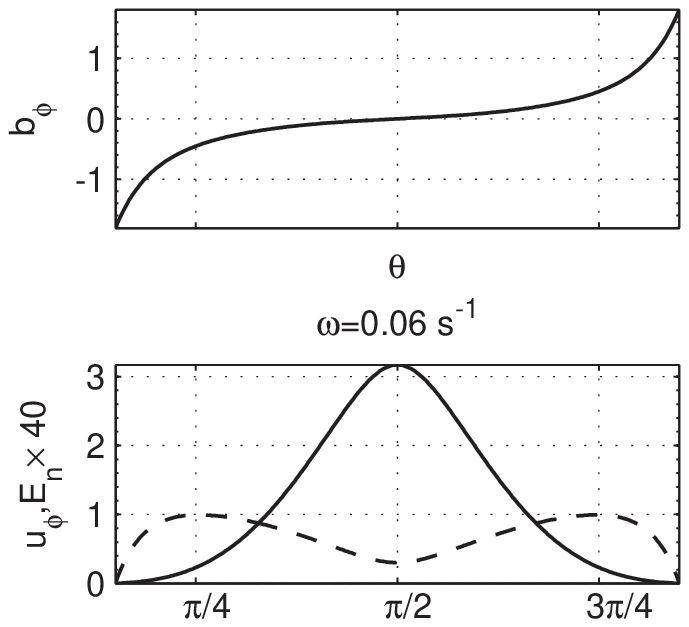}
\\\hspace{.07\textwidth} (a) \hspace{.41\textwidth} (b)
\end{center}
\caption{Wave forms of the fundamental mode ($L=6$) on the (a)
noon
 and (b) midnight  meridional plane ($\phi=0$ and $\pi$),
respectively. Dashed line denotes the electric field profile which
is multiplied by 40. All units are arbitrary. The corresponding
eigen-frequency  is denoted in the middle of each
subplot.}\label{fig2}
\end{figure}

\begin{figure*}[htb]
\begin{center}
\includegraphics[width=12cm]{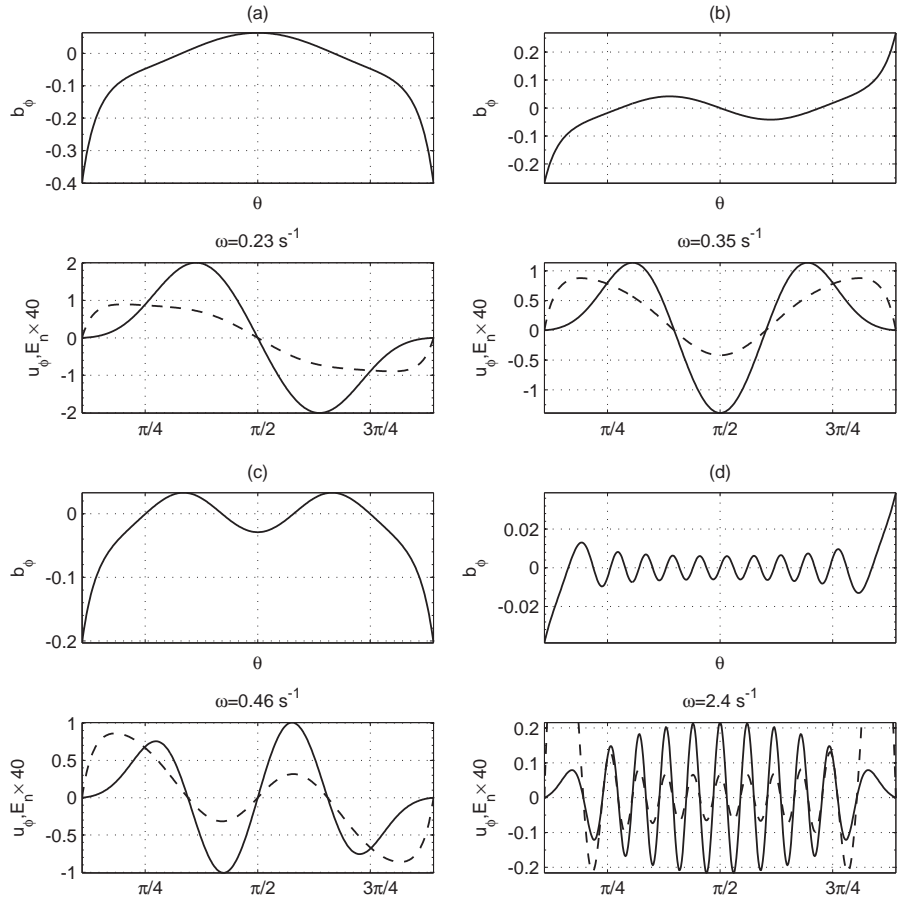}
\end{center}
\caption{Harmonic wave forms (arbitrary unit) as derived from the
solutions of the KG equation on the noon meridional plane
($\phi=0$) of the compressed dipole field for (a) n=1, (b) n=2,
(c) n=3, and (d) n$>$4. Format is the same as Fig.~\ref{fig2}.
}\label{fig4}
\end{figure*}

\begin{figure*}[htb]
\begin{center}
\includegraphics[width=12cm]{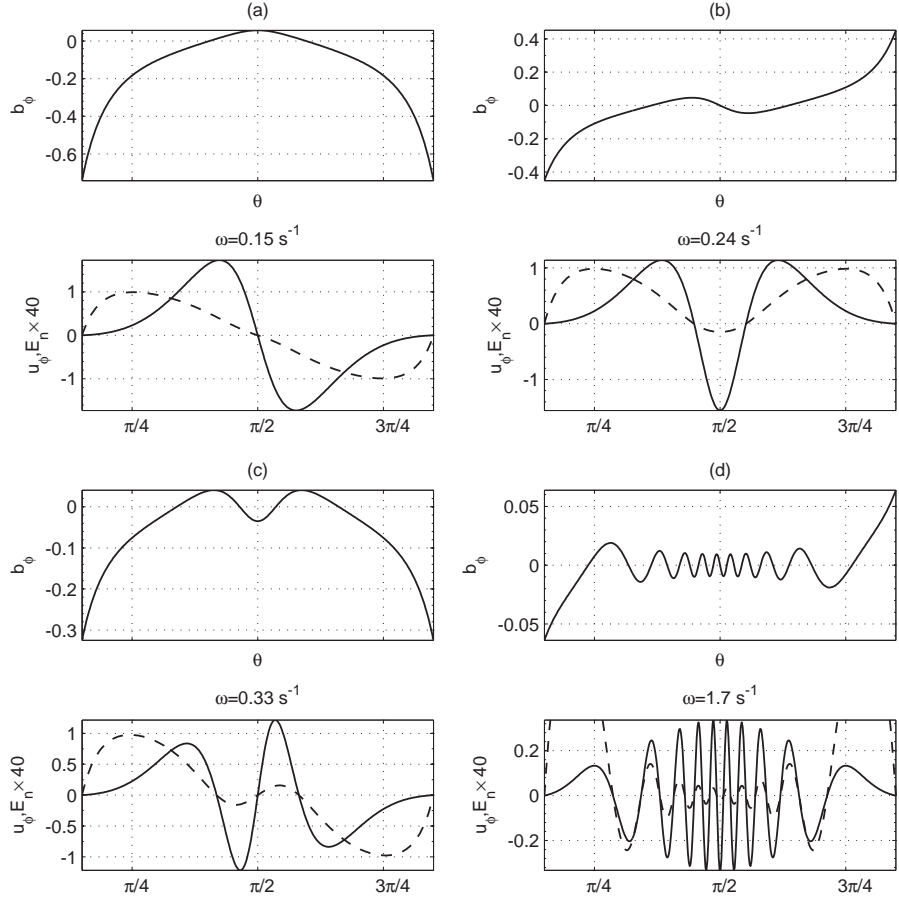}
\end{center}
\caption{Harmonic wave forms on the midnight meridional plane
($\phi=\pi$) of the compressed dipole field. Format is the same as
Fig.~\ref{fig4}.}\label{fig5}
\end{figure*}

The choice of  $L=6$, which shows the greatest asymmetry between
the noon side and midnight side of the compressed dipole, is a
representative case to illustrate the spatial wave forms as
harmonic solutions to the KG equation. The number of nodes, n,
contained in the solution of $u_\phi$ is increasing from 0 in the
fundamental mode to consecutive positive integral numbers for
higher-order harmonics. Fig.~\ref{fig2}a, b show the fundamental
modes for the noon and midnight side meridional planes of the
compressed dipole. Similar to a standard dipole case
\citep{2012JGRAW}, the $b_\phi$ profile contains one node at the
equator, and the oscillating velocity $u_\phi$ and electric field
$E_n$, normal to the background field (see Fig.~\ref{fig0}) are in
phase, given the boundary condition $E_n=0$ at both footpoints.
The fundamental mode frequency at the midnight side is a little
smaller than the noon side and the corresponding $E_n$ profile has
a significant dip (much reduced amplitude) near the equator. These
differences are caused by the different field-line geometry, the
cut-off frequency and the amplitude factor for the two sides as
discussed earlier. Such differences persist for higher-order
harmonics. Figs.~\ref{fig4} and \ref{fig5} show the wave forms of
higher-order harmonics of increasing number of nodes on the noon
and midnight side, respectively. The eigen-frequency increases
with increasing number of nodes. The $u_\phi$ and $E_n$
perturbations remain in phase while the $b_\phi$ oscillation is
generally out of phase by $\pi/2$. For the same harmonic mode, the
midnight-side solution always has a smaller eigen-frequency and a
smaller amplitude in $E_n$ around the equator.

\section{Eigenvalue Solutions of the Poloidal Mode} \label{pol}
In the poloidal mode, both the magnetic field and velocity
perturbations of the waves are in the meridional plane. The normal
components perpendicular to the field line (see Fig.~\ref{fig0})
are denoted $b_n$ and $u_n$, respectively. Therefore, the only
oscillating electric field is along the $\hat\phi$ direction,
$E_\phi$, and after multiplied by  a scaling factor,
$\epsilon_\phi=r\sin\theta E_\phi$, is governed by
\citep{1969JGRC,1993AandAO,2006GMSL}
\begin{equation}
\nabla^2 \epsilon_\phi +
2r\sin\theta\nabla\epsilon_\phi\cdot\nabla\left(\frac{1}{r\sin\theta}\right)
+\frac{\omega^2}{V_A^2}\epsilon_\phi=0, \label{eq:Ephi}
\end{equation}
again assuming a harmonic time variation with angular
eigen-frequency, $\omega$.  The scaling factor $r\sin\theta$
arises from the curvilinear coordinate system other than a
Cartesian geometry. In an equivalent cylindrical coordinate
$(R,\phi, Z)$ ($\partial/\partial \phi=0$), it is written
\begin{equation}
 \frac{\partial^2\epsilon_\phi}{\partial R^2} -
\frac{1}{R}\frac{\partial\epsilon_\phi}{\partial
R}+\frac{\partial^2\epsilon_\phi}{\partial Z^2}
+\frac{\omega^2}{V_A^2}\epsilon_\phi=0. \label{eq:Ephi2}
\end{equation} In a
Cartesian geometry, the differential operator in the above
equation becomes a single Laplacian  and $\epsilon_\phi\equiv
E_\phi$ \citep[e.g.,][]{1969JGRC}. Here the Alfv\'{e}n speed
$V_A=B/\sqrt{\mu_0\rho}$ is again determined by a given background
magnetic field and density model, and the equation is solved in a
2D domain such as a meridional plane of the compressed dipole
field for $\phi=0$ and $\pi$ only. We are seeking eigen-mode
solutions subject to boundary condition $\epsilon_\phi\equiv 0$ in
the present study. Once the electric field perturbation is
obtained, the magnetic field perturbations, $b_r$ and $b_\theta$,
lying in the meridional plane, can be derived via Faraday's law
using the linear approximations.

Interestingly, the magnetic field perturbation normal to the field
line, $b_n$, can be derived along each individual field line
following the previous approach by the equation below which
follows from the linearized Faraday's law:
\begin{equation}
\frac{\partial{b_n}}{\partial
t}=-\frac{B_\theta}{r^2B\sin\theta}\frac{d\epsilon_\phi}{d\theta}.
\label{eq:bn}
\end{equation}
Note that the total derivative $d/d\theta$ here is evaluated along
each individual field line and takes the form of Eq.~(\ref{ddth}).
 For harmonic oscillations, if we assign a phase lag of $\pi/2$ to $b_n$ relative to $E_\phi$ at initial time,  the left-hand side of Eq.~(\ref{eq:bn}) becomes
  $\omega \tilde{b}_n$, which allows the derivation of a real-valued amplitude profile of $b_n$ based on solutions to Eq.~(\ref{eq:Ephi}).
Similarly, the tangential component of the magnetic field
perturbation is obtained by
\begin{equation}
\frac{\partial b_s}{\partial
t}=\frac{1}{r\sin\theta}(\nabla\epsilon_\phi\cdot{\hat
n})=\frac{1}{r\sin\theta}\left(-\frac{B_r}{rB}\frac{\partial\epsilon_\phi}{\partial\theta}+\frac{B_\theta}{B}\frac{\partial\epsilon_\phi}{\partial
r}\right).\label{eq:bs}
\end{equation}
In general, the right-hand side of the above equation does not
vanish, indicating a compressional fast mode solution. On the
other hand, if it does vanish, i.e.,
$\partial\epsilon_\phi/\partial n=0$, a standing Alfv\'{e}n wave
mode should result. We separately analyze these two wave modes in
the following subsections.


\subsection{Poloidal Standing Transverse (Alfv\'{e}n) Mode}
This is a special case corresponding to $b_s\equiv 0$, i.e.,
$\partial\epsilon_\phi/\partial n=0$ from Eq.~(\ref{eq:bs}) above.
This corresponds to a transverse, Alfv\'{e}n mode of poloidal
polarization of the magnetic field perturbation that is
propagating along individual field lines. Therefore we can apply
exactly the same approach of Section~\ref{gov}. The electric field
perturbation $\epsilon_\phi$ still satisfies Eq.~(\ref{eq:Ephi}).
However when applying the condition $b_s=0$ and projecting the PDE
along an individual field line defined by a relation between $r$
and $\theta$, a wave equation of the form similar to
Eq.~(\ref{eqKG}) is obtained
\begin{equation}
\frac{\partial^2\epsilon_\phi}{\partial
t^2}=\frac{V^2}{r^2}\left[\frac{d^2\epsilon_\phi}{d\theta^2}-\frac{d}{d\theta}\ln\left(\frac{B^2}{B_\theta^2}g(\theta)\sin\theta\right)\frac{d\epsilon_\phi}{d\theta}\right].\label{eq:KGpol}
\end{equation}
Here the wave speed parameter $V^2\equiv B^2_\theta/(\mu_0\rho)$
remains the same as before, and the function $g(\theta)$ is
determined from a given background magnetic field model along an
individual field line $r(\theta)$ by
\begin{equation}
\frac{d}{d\theta}\ln
g(\theta)=\frac{B_r^2}{B_\theta^2}\frac{\partial}{\partial
r}\left(\frac{rB_\theta}{B_r}\right).\label{eq:gpol}
\end{equation}
For example, for a standard dipole field, the function
$g(\theta)=\sin^2\theta$ is obtained.

\begin{figure}[htb]
\begin{center}
\includegraphics[width=8.3cm]{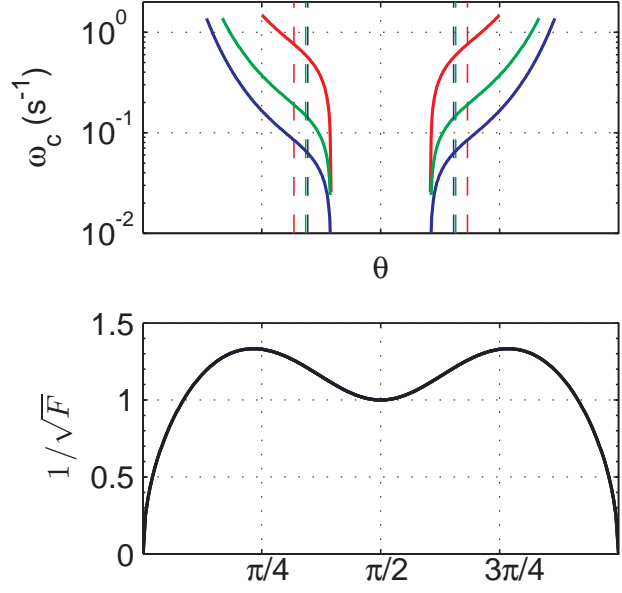}
\end{center}
\caption{The parameters $\omega_c$ and $f(\theta)$ for the
poloidal standing Alfv\'{e}n waves of the dipole field. Format is
the same as Fig.~\ref{fig00}. }\label{fig00pol}
\end{figure}

\begin{figure}[htb]
\begin{center}
\includegraphics[width=.45\textwidth]{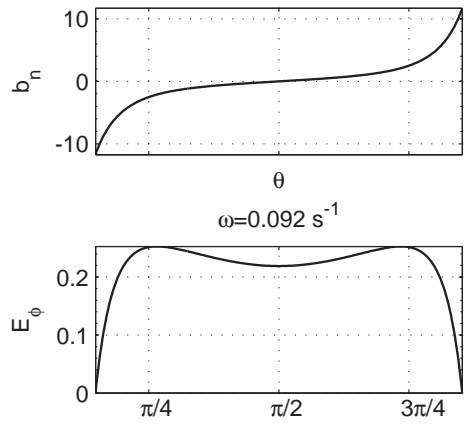}
\end{center}
\caption{Wave forms (arbitrary unit) of the fundamental mode
($L=5$) for the poloidal standing Alfv\'{e}n mode of the dipole
field. Format is the same as Fig.~\ref{fig2}. }\label{fig2pol}
\end{figure}

Therefore the wave equation~(\ref{eq:KGpol}) can also be cast into
a KG form and solved for eigenvalue solutions subject to the
boundary condition $\epsilon_\phi=0$  at the footpoints of an
individual field line. In turn the magnetic field perturbation can
be derived from Eq.~(\ref{eq:bn}). Below we list the essential
parameters for this mode conforming to the general descriptions in
Section~\ref{sec:general}:
\begin{eqnarray}
\frac{1}{L_\epsilon}&=&\frac{d}{d\theta}
\ln\left(\frac{B^2}{B_\theta^2}g(\theta)\sin\theta\right)\\
\frac{1}{M_\epsilon}&=&0,
\end{eqnarray}
and the amplitude factor
\begin{equation}
f(\theta)=\frac{B}{B_\theta}\sqrt{g(\theta)\sin\theta}.
\end{equation}
For the dipole field, the following explicit formulas are obtained
\begin{equation}
f(\theta)=\sqrt{\sin\theta(1+3\cos^2\theta)}\label{eq:f_dpl_pol}
\end{equation}
\begin{equation}
\frac{1}{L_\epsilon}=\cot\theta-\frac{3\sin
2\theta}{1+3\cos^2\theta}.\label{eq:Lpol}
\end{equation}
Thus  the cut-off frequency $\omega_c$ can be written based on
Eq.~(\ref{eqwc2}). Fig.~\ref{fig00pol} shows, in the same format
as before, the profiles of $\omega_c$ and $f$ for selected $L$
shells of the dipole field. Similarly the cut-off frequency
exhibits minimum near the equator where $\omega_c^2$ becomes
negative and increases toward the poles. The eigen-frequency of
the fundamental mode generally intersects the cut-off frequency at
low to mid latitudes. The solution of the KG equation would also
be a combination of a propagation type near the equator and a
decaying type near the two ends. The amplitude factor $f(\theta)$,
on the other hand, shows much less variation in magnitude and does
not depend on $L$. Fig.~\ref{fig2pol} shows the fundamental mode
solutions for $L=5$, typical of a standing wave with zero number
of node in electric field perturbation. The amplitude of $E_\phi$
dips slightly around the equator. The eigen-frequency is 0.092
s$^{-1}$, which corresponds to a period of 68s. It compares well
with observations
 to be discussed below. Table~\ref{tbl:pol} lists the
corresponding parameters for the selected $L$ shells in the same
format as Tables~\ref{tbl:dpl}-\ref{tbl:cdpl1}. The periods are in
the same range as those of the toroidal mode.

\begin{figure}[htb]
\centering
\includegraphics[width=8.3cm]{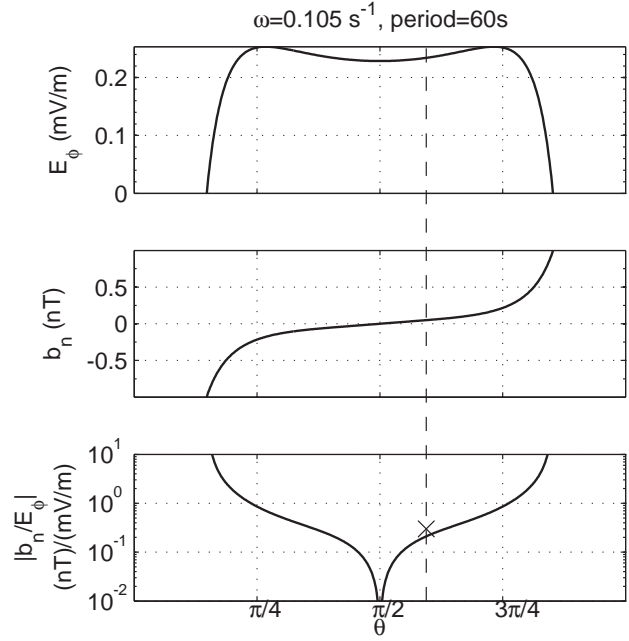}
\caption{The solutions of $E_\phi$ (in mV/m), $b_n$ (in nT), and
the ratio in-between (from top to bottom panels), to be compared
with the spacecraft observations of \citet{2013GeoRLD}. The
vertical dashed line denotes the spacecraft location and the cross
sign marks the measured ratio ($\sim 0.3$ nT/(mV/m)) during the
time period of measurements.}\label{fig:Dai}
\end{figure}

For a compressed dipole field, because the relation below $r$ and
$\theta$ along a field line is implicit (see the Appendix),  the
relevant quantities have to be evaluated numerically. We leave
detailed analysis of this particular case for future study. In
what follows, we demonstrate the validity of our approach by
comparing it with a most recent direct spacecraft observation of
poloidal standing Alfv\'{e}n waves by \citet{2013GeoRLD}.
Fig.~\ref{fig:Dai} shows our results of suitable physical units
for $L=5$ of the dipole field with the same set of parameters as
\citet{2013GeoRLD}, $\rho_e=6.4$ amu cm$^{-3}$ and the power index
1.0 of the density variation, to facilitate a direct comparison
with their results (Fig. 3 therein). \citet{2013GeoRLD} used a
theoretical model of \citet{1969JGRC} and realistic ionosphere
boundary conditions of finite  conductivity at the footpoints of
the field line. Therefore their solutions of $E_\phi$ and $b_n$
profiles are of finite values at the ends and are asymmetric about
the equator whereas ours are symmetric and $E_\phi$ vanishes at
the two ends. Nonetheless the spatial profiles over the low-mid
latitudes still compare very well. The magnitudes of both
perturbations show a slight decrease toward the equator in
$E_\phi$ and rapid increase toward the ends in $b_n$. In
particular, the ratio of $|b_n/E_\phi|$ at the spacecraft location
($\sim 17^\circ$ south in latitude),  $\sim$ 0.21 nT/(mV/m),
agrees well with their result, 0.25 nT/(mV/m), and the observed
value, 0.3 nT/(mV/m). Similarly the period of the wave, 60s,
compares well with 62s by \citet{2013GeoRLD} and the observed
value, 84s. Our result is also consistent with the observation in
that $E_\phi$ leads the phase of $b_n$ by $\pi/2$ as discussed
earlier.

\subsection{Poloidal Compressional Mode}
\begin{figure}[htb]
\centering
\includegraphics[width=8.3cm]{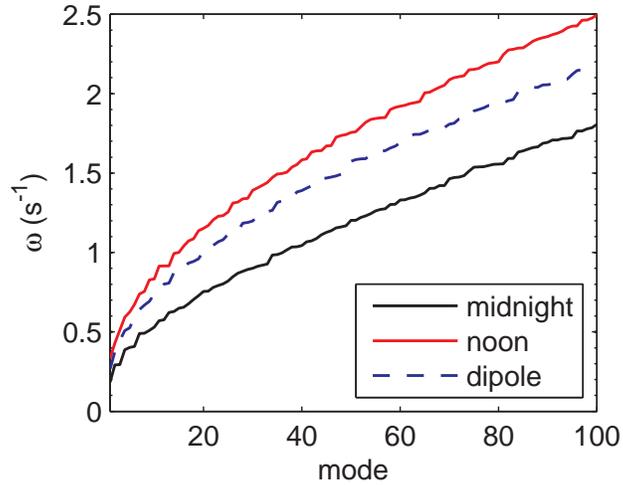}
\caption{The first 100 eigen-frequencies for the noon side and
midnight side of the compressed dipole, and a standard dipole
field. }\label{fig:omgs}
\end{figure}

In the case that $b_s\ne 0$, Eq.~(\ref{eq:Ephi2}) has to be solved
in a two-dimensional domain as an eigenvalue problem subject to
the boundary condition $E_\phi=0$ on all sides. As illustrative
examples, we solve the equation and obtain the corresponding
eigen-mode solutions of a discrete set of increasing
eigen-frequencies by utilizing the software package
PDE2D\footnote{\texttt{http://www.pde2d.com/}}. The solutions are
also cross-checked with the Matlab PDE  toolbox and identical
results are obtained. The computational domain is chosen as
$r\in[1, L_p]R_E$, and $\theta\in[\theta_p, \pi-\theta_p]$, where
$\theta_p=\arcsin\sqrt{1/L_p}$. We choose $L_p=7$ in order to
avoid the singular point in the compressed dipole field model as
well as the singularity along the poles (X=0). We apply the dipole
field and the compressed dipole field models and the same density
distribution as before for the background field and plasma
conditions. Three sets of eigen-frequencies of ascending order of
magnitude (mode) are obtained for both the noon and midnight side
meridional planes of the compressed dipole field and the standard
dipole field. The first 100 eigen-frequencies are shown in
Fig.~\ref{fig:omgs}. They generally exhibit a rapid rise at the
lowest numbers of mode, then the trend of increase seems to become
more gradual and eventually linear. At one specific mode, the
eigen-frequency of the noon side is always greater than that of
the midnight side, while the value of the dipole case is always
in-between, albeit slightly closer to the value of the noon side.

\begin{figure*}[htb]
\centering
\includegraphics[width=.324\textwidth]{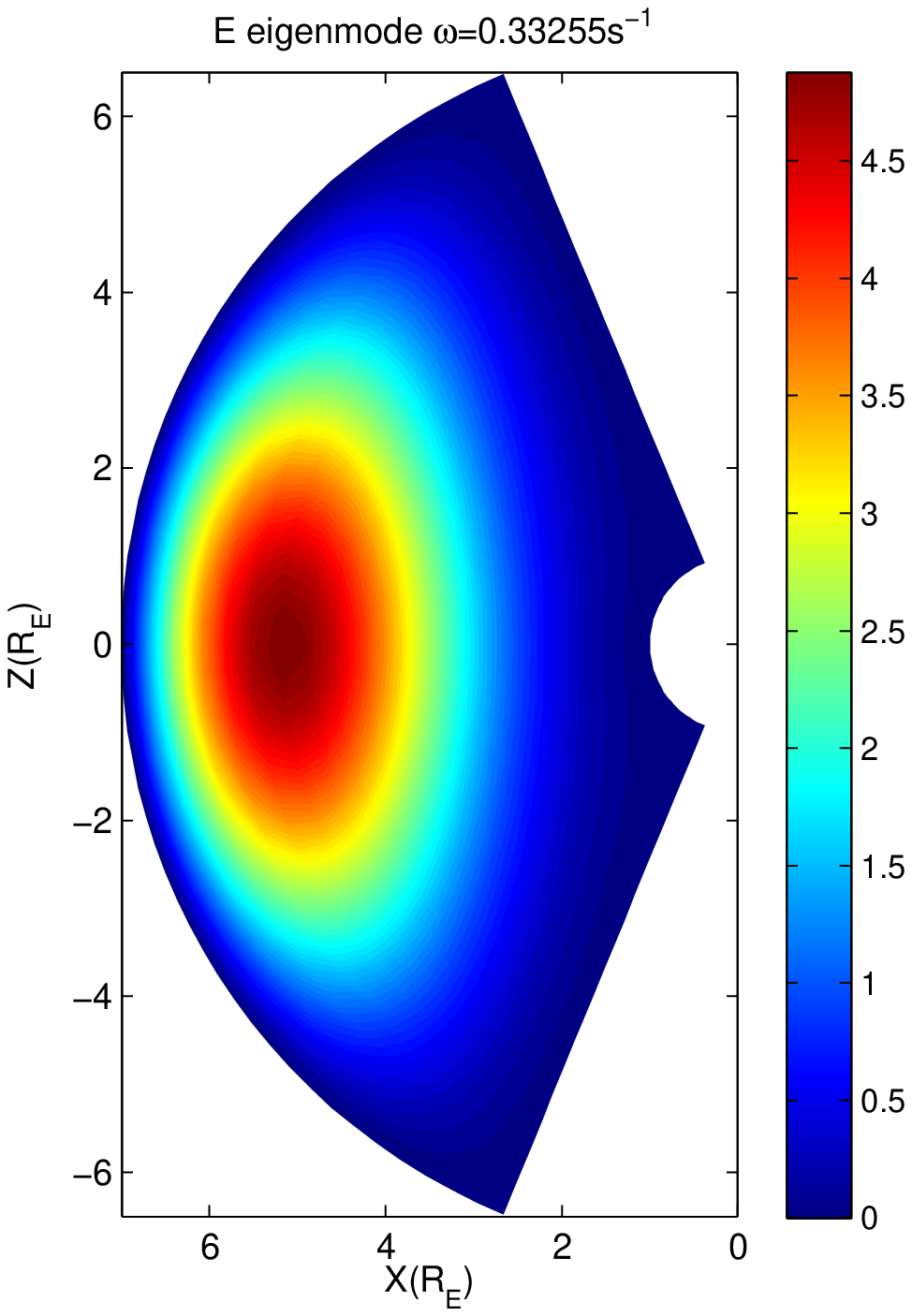}
\includegraphics[width=.329\textwidth]{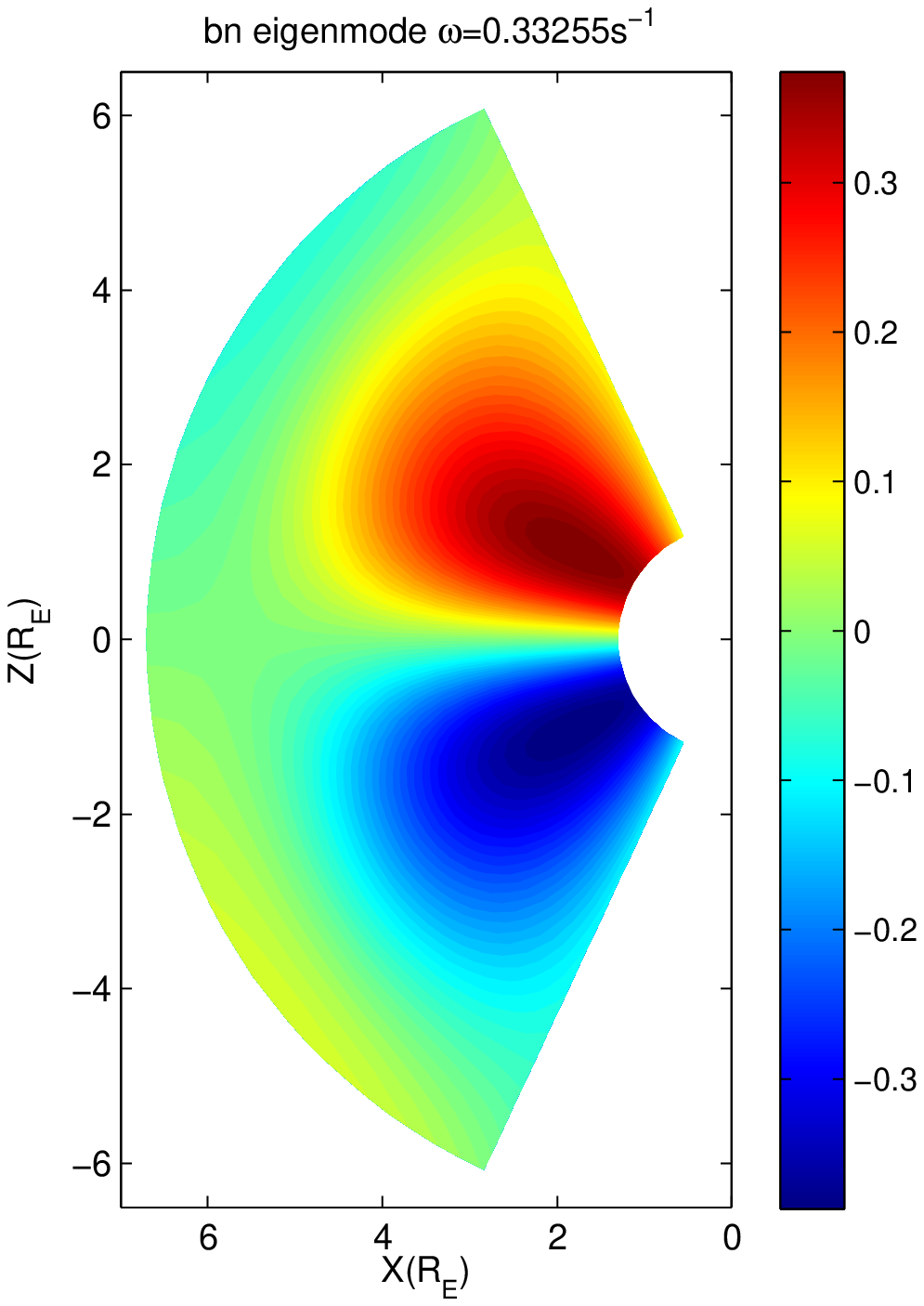}
\includegraphics[width=.329\textwidth]{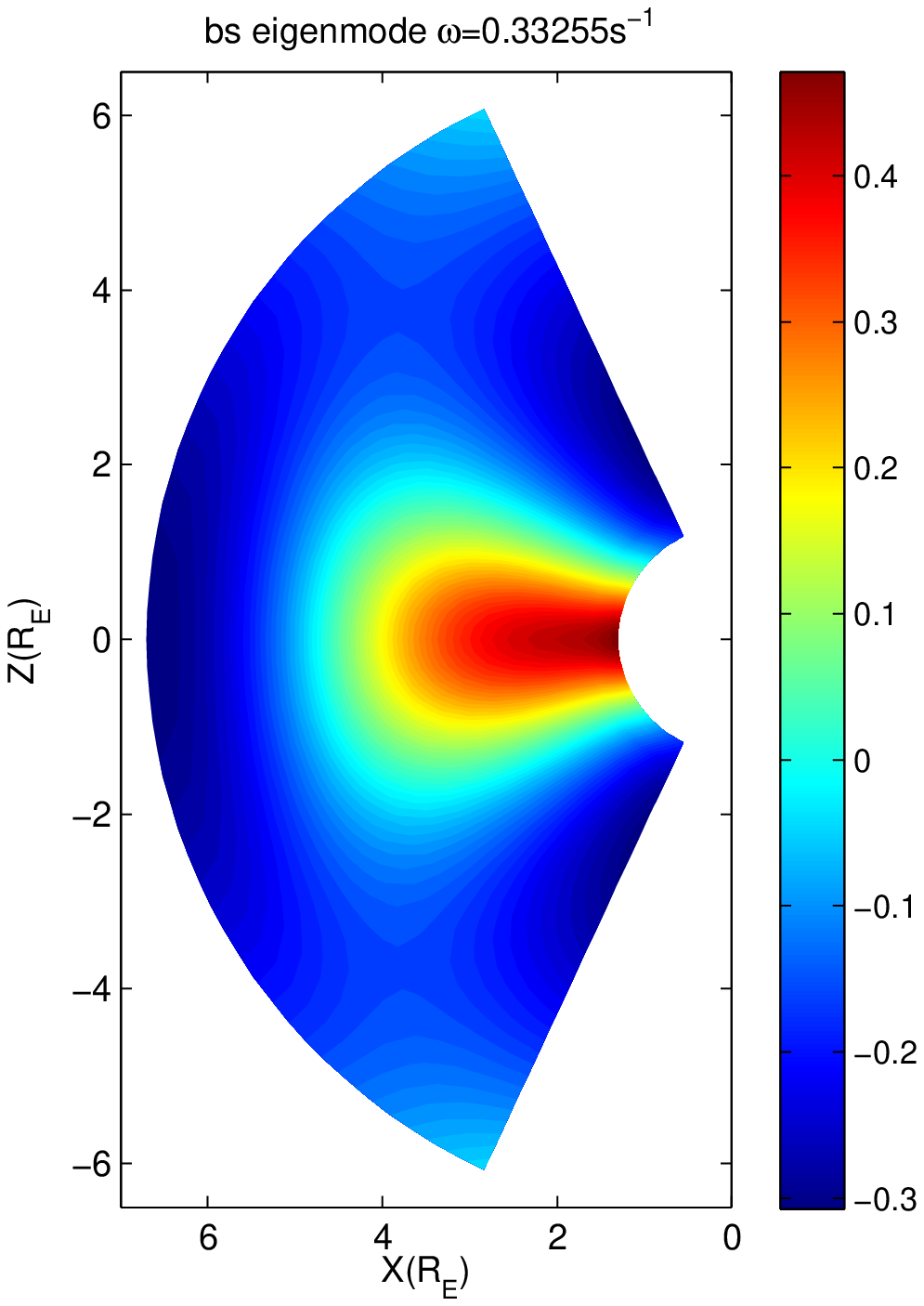}\\
\includegraphics[width=.323\textwidth]{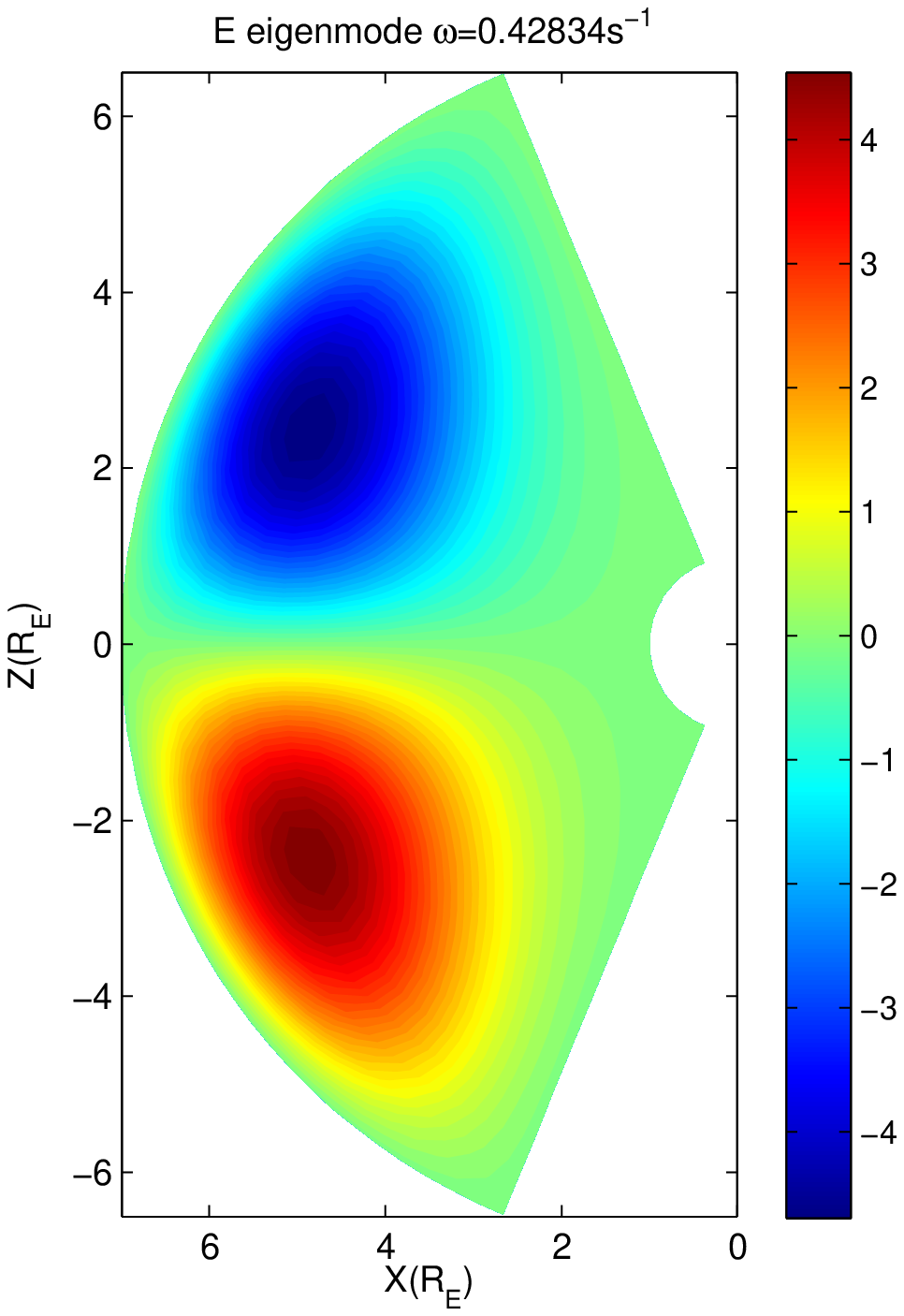}
\includegraphics[width=.329\textwidth]{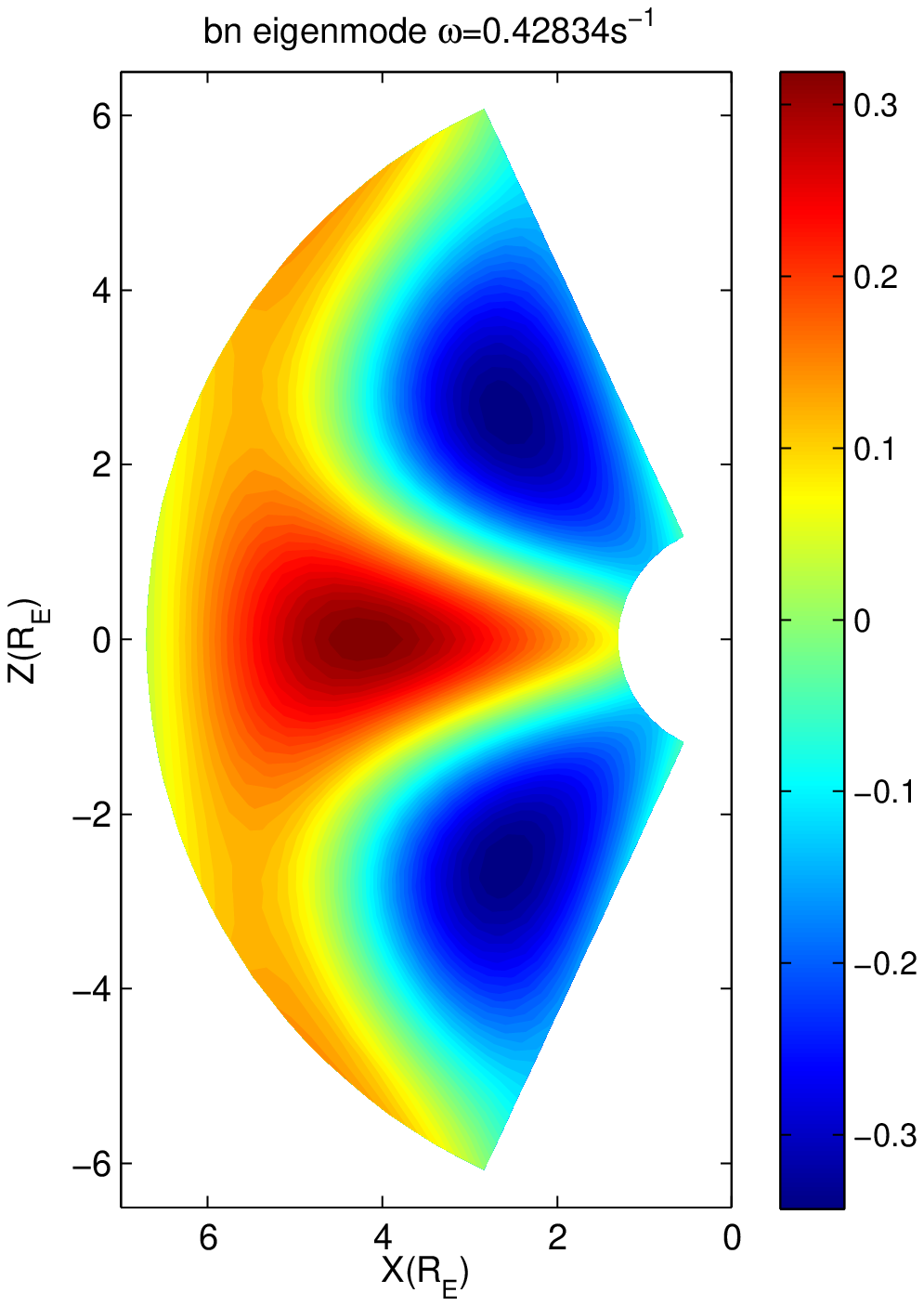}
\includegraphics[width=.329\textwidth]{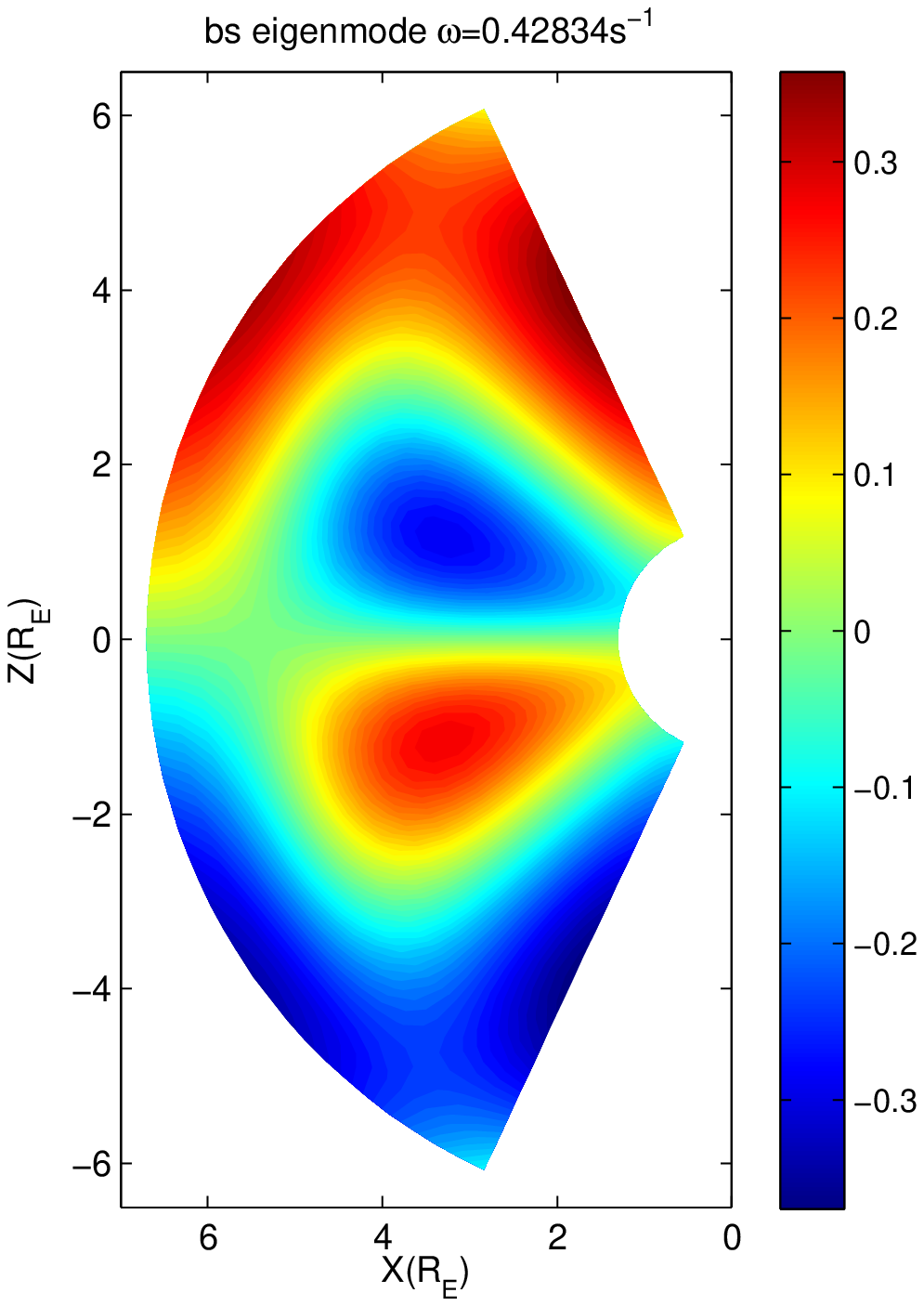}\\
\caption{The fundamental (top row) and 1st-order harmonic (bottom
row) eigen-mode solutions of $E_\phi$, $b_n$, and $b_s$ (from left
to right columns) for $\phi=0$ (noon side) of the compressed
dipole field. The corresponding eigen-frequencies are indicated on
top of each panel. The scales (all of arbitrary units) are given
by the colorbars. }\label{fignoon}
\end{figure*}

\begin{figure*}[htb]
\centering
\includegraphics[width=.327\textwidth]{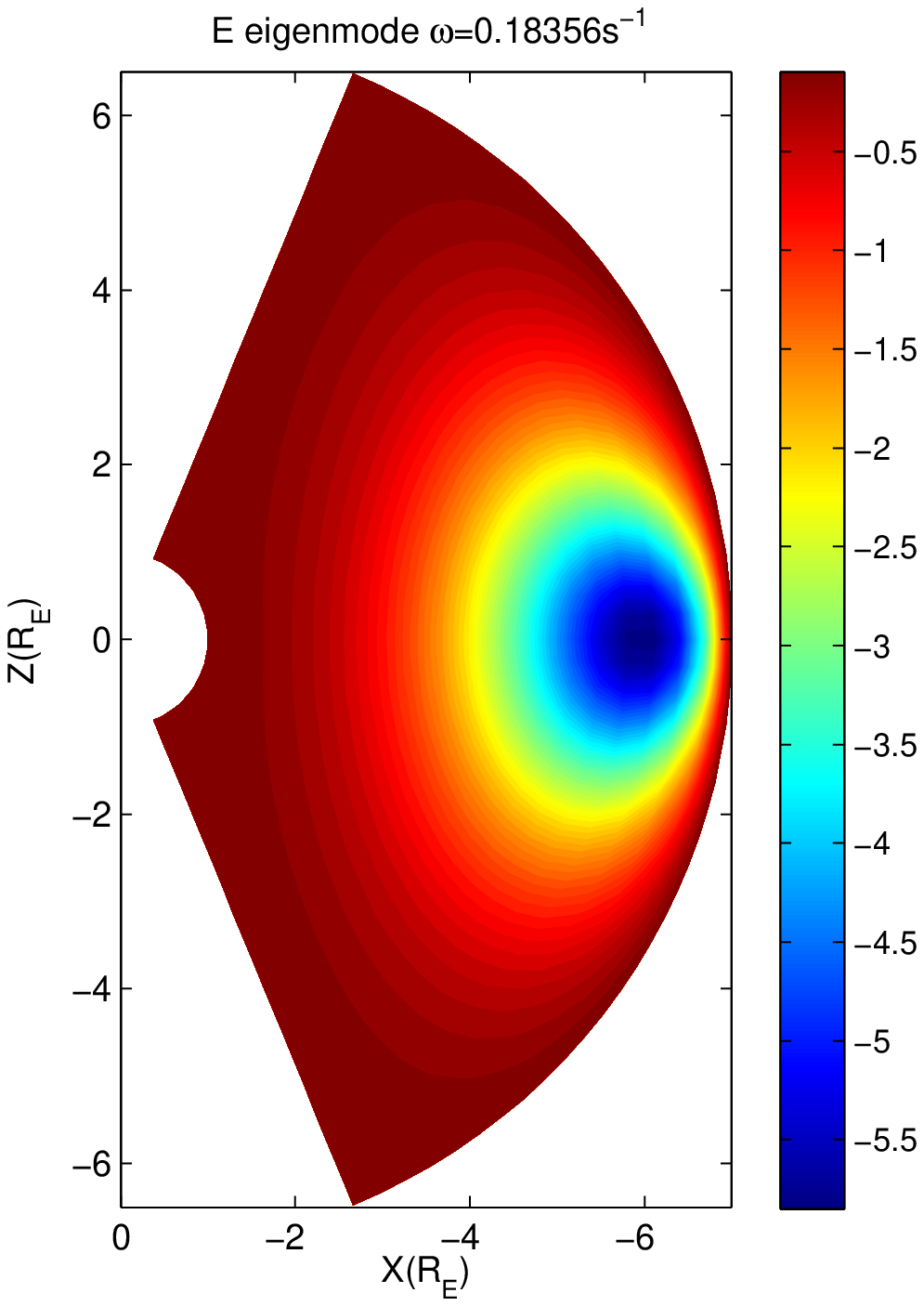}
\includegraphics[width=.328\textwidth]{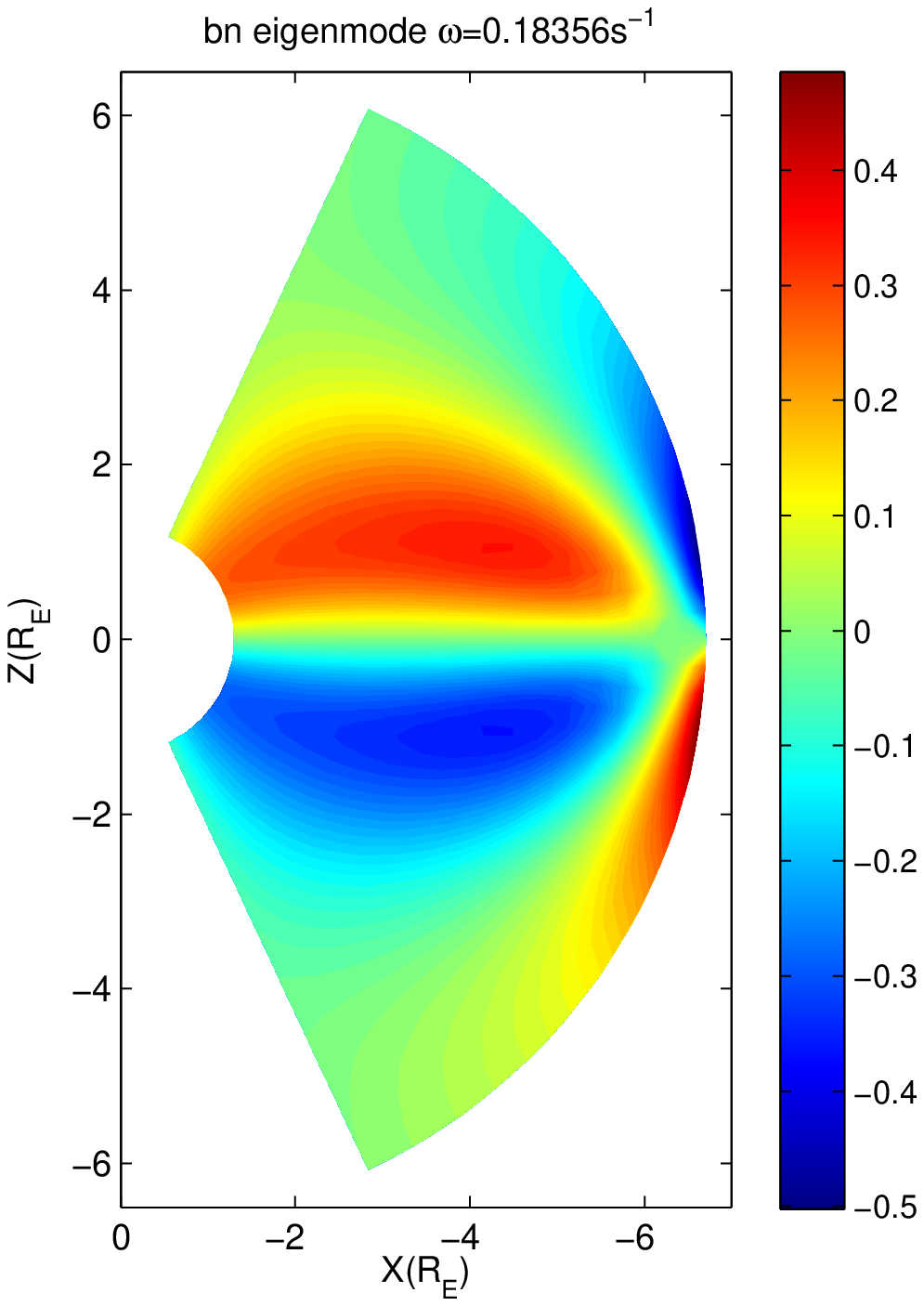}
\includegraphics[width=.327\textwidth]{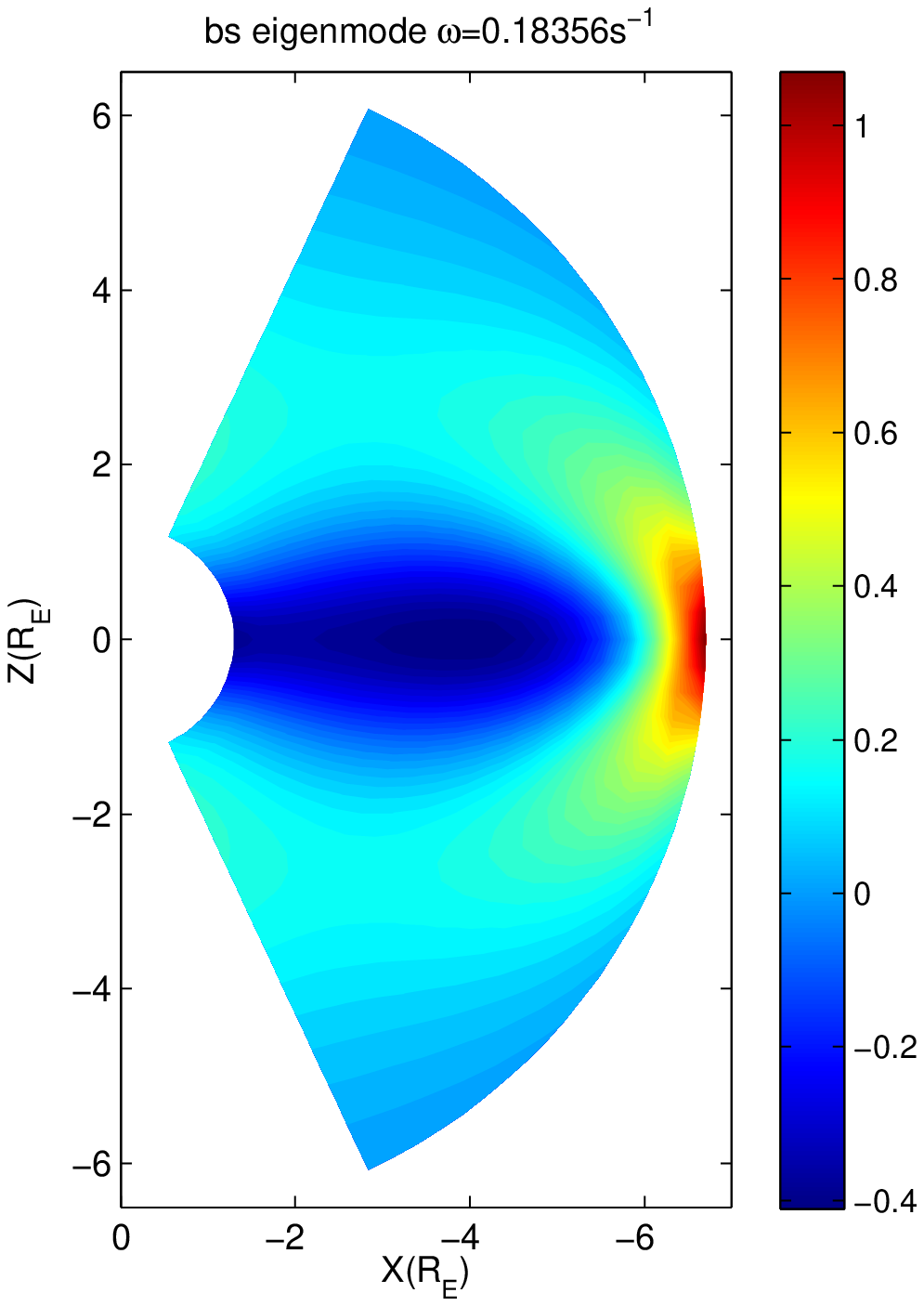}\\
\includegraphics[width=.323\textwidth]{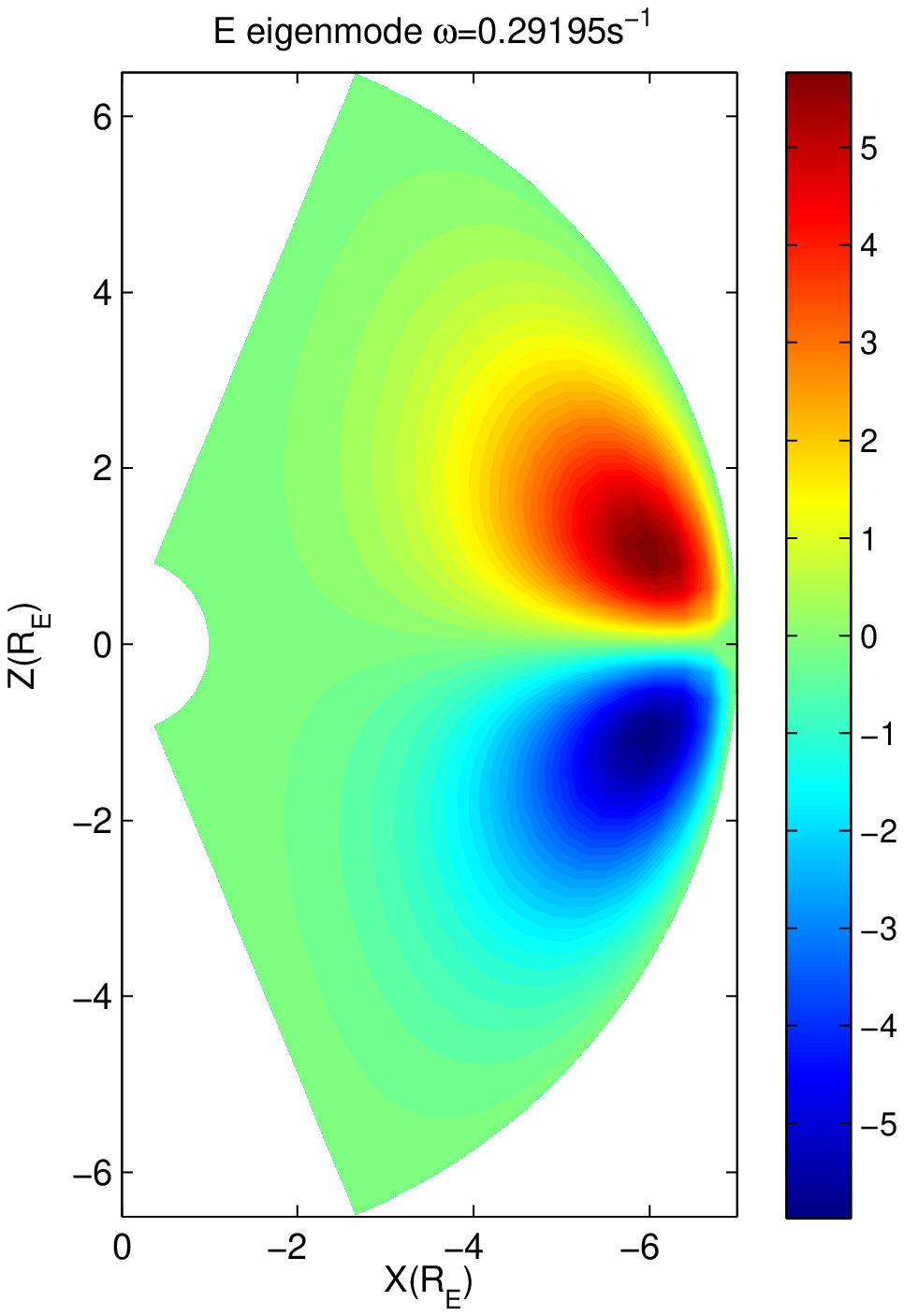}
\includegraphics[width=.33\textwidth]{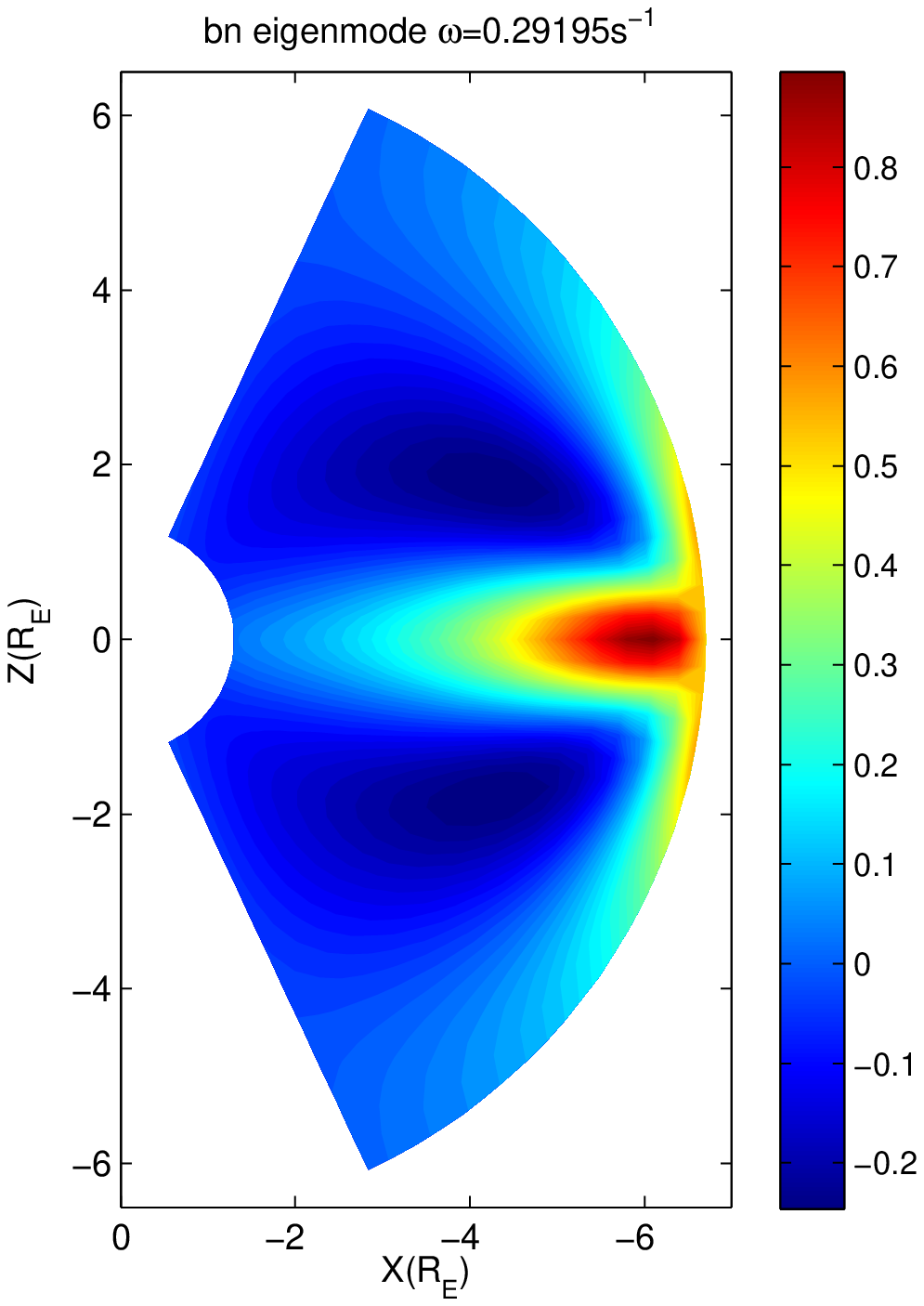}
\includegraphics[width=.327\textwidth]{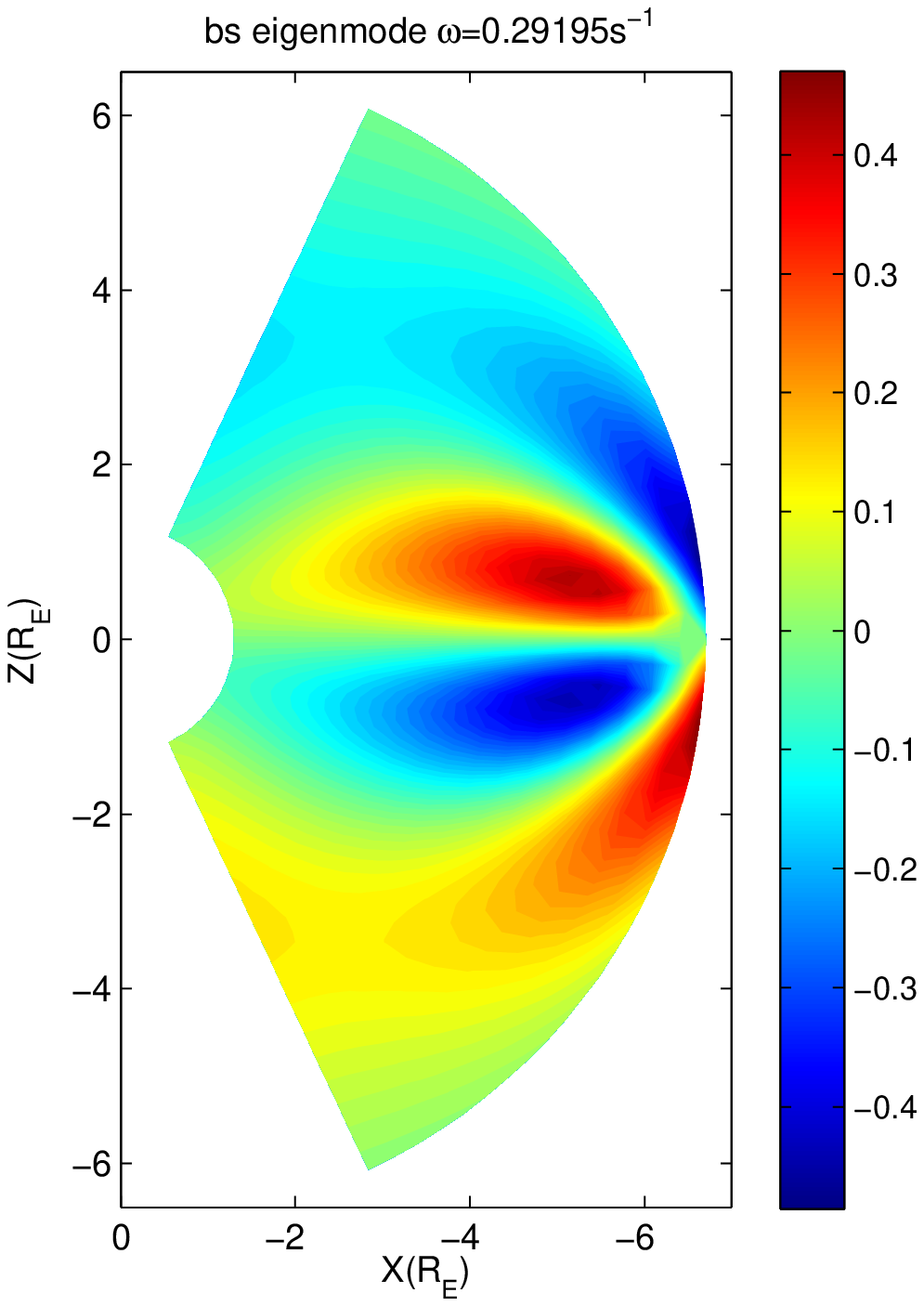}\\
\caption{The eigen-mode solutions for $\phi=\pi$ (midnight side)
of the compressed dipole field. Format is the same as
Fig.~\ref{fignoon}.}\label{figmidnight}
\end{figure*}

\begin{figure*}[htb]
\begin{center}
\includegraphics[width=.328\textwidth]{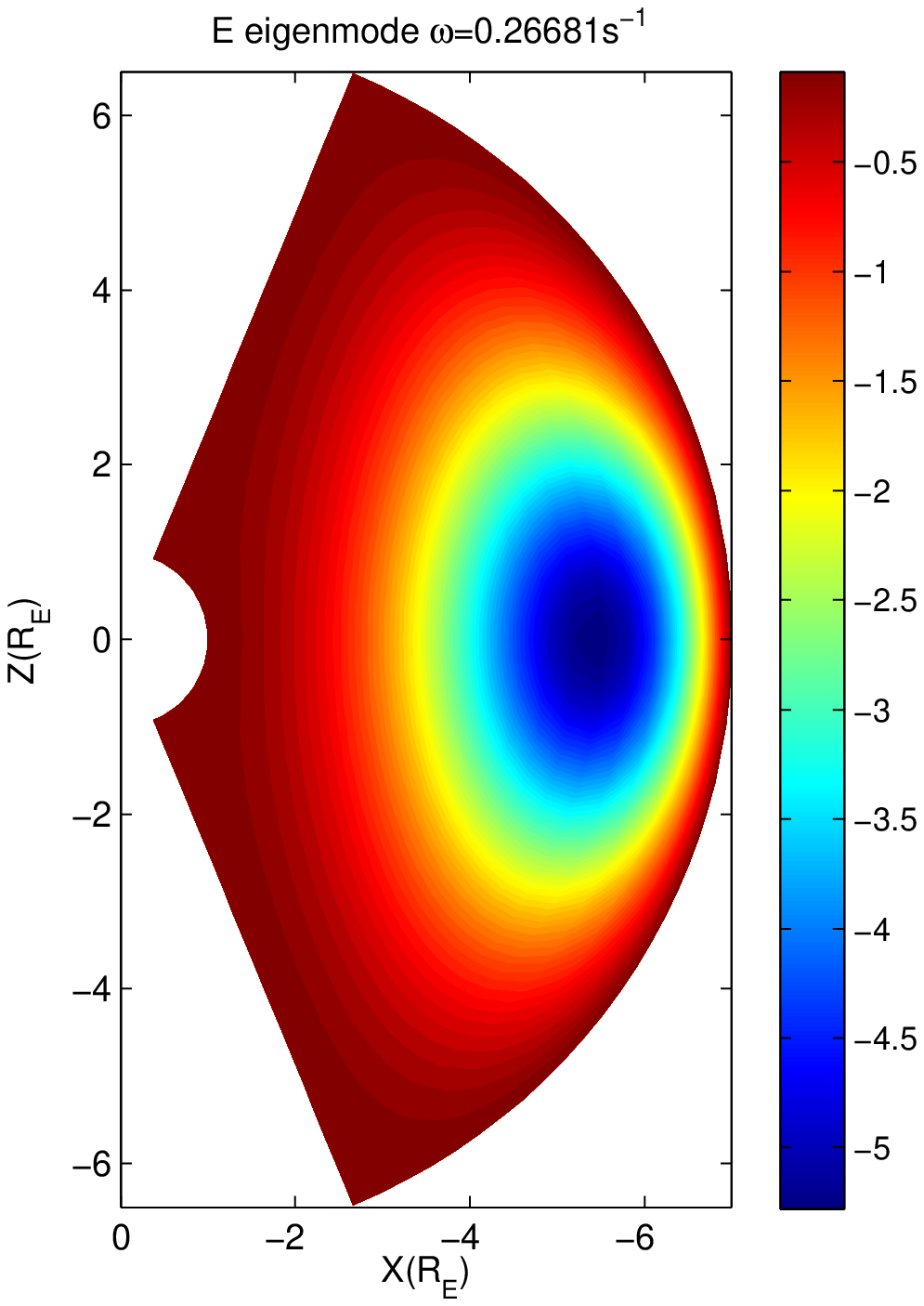}
\includegraphics[width=.327\textwidth]{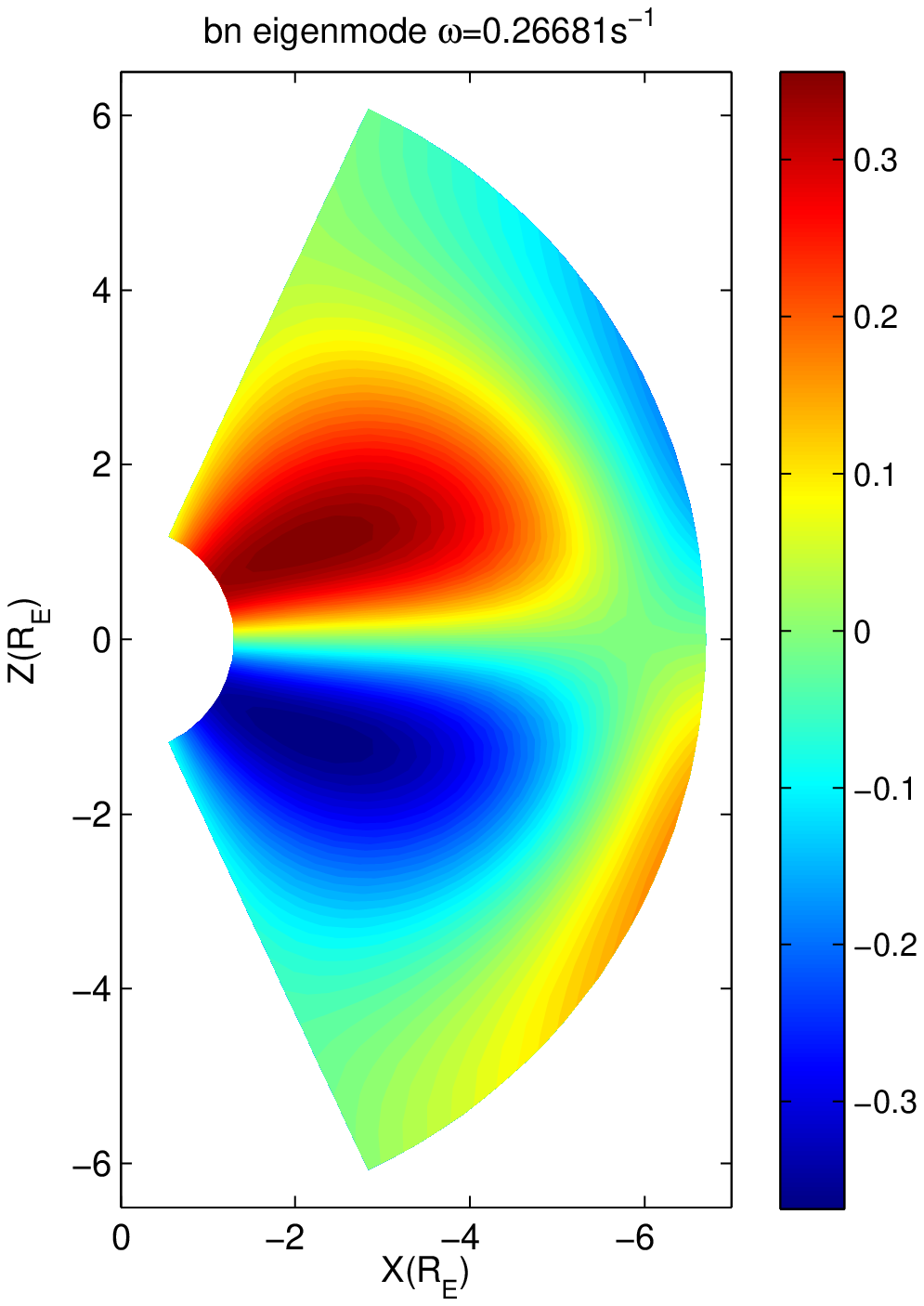}
\includegraphics[width=.327\textwidth]{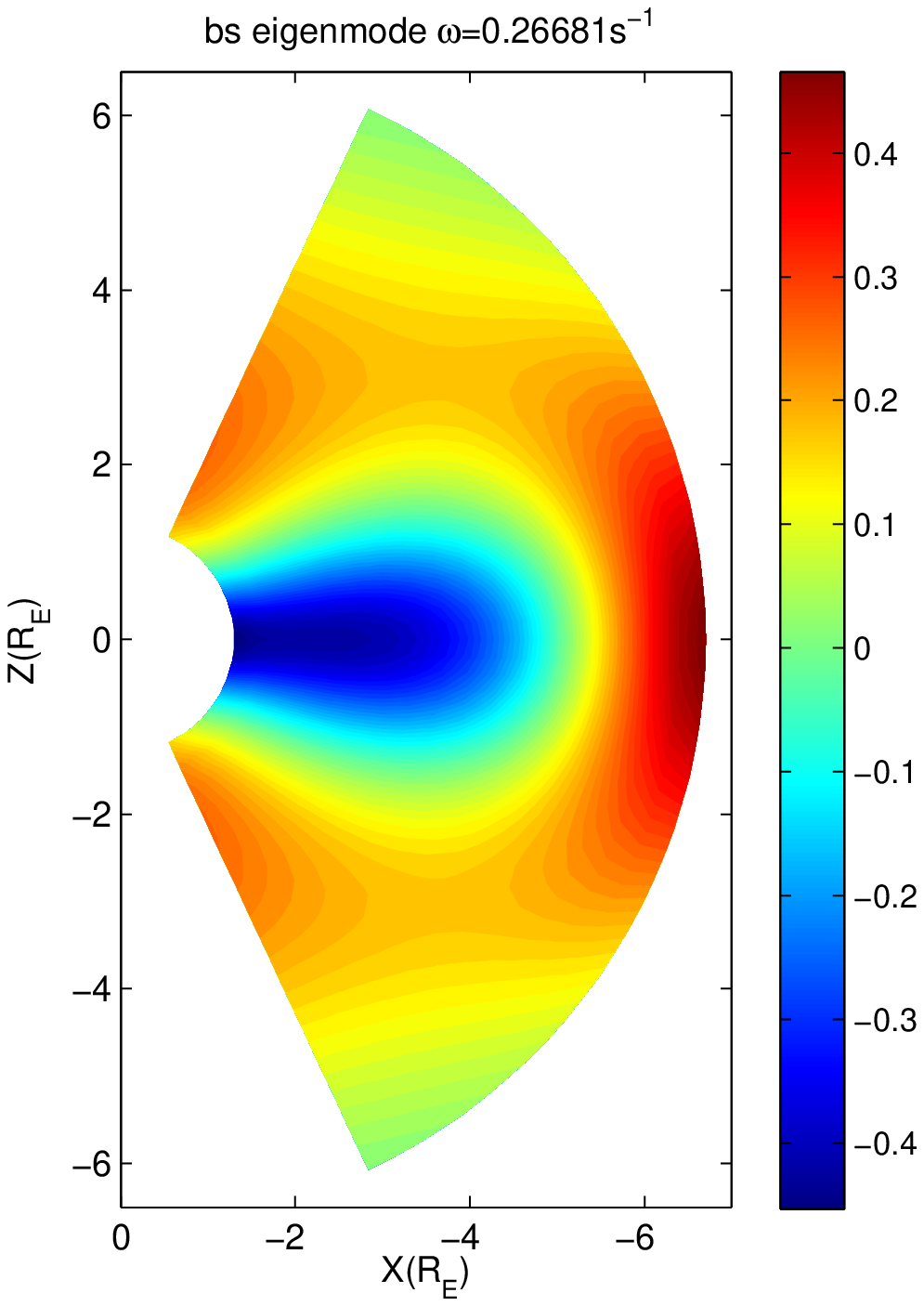}\\
\includegraphics[width=.324\textwidth]{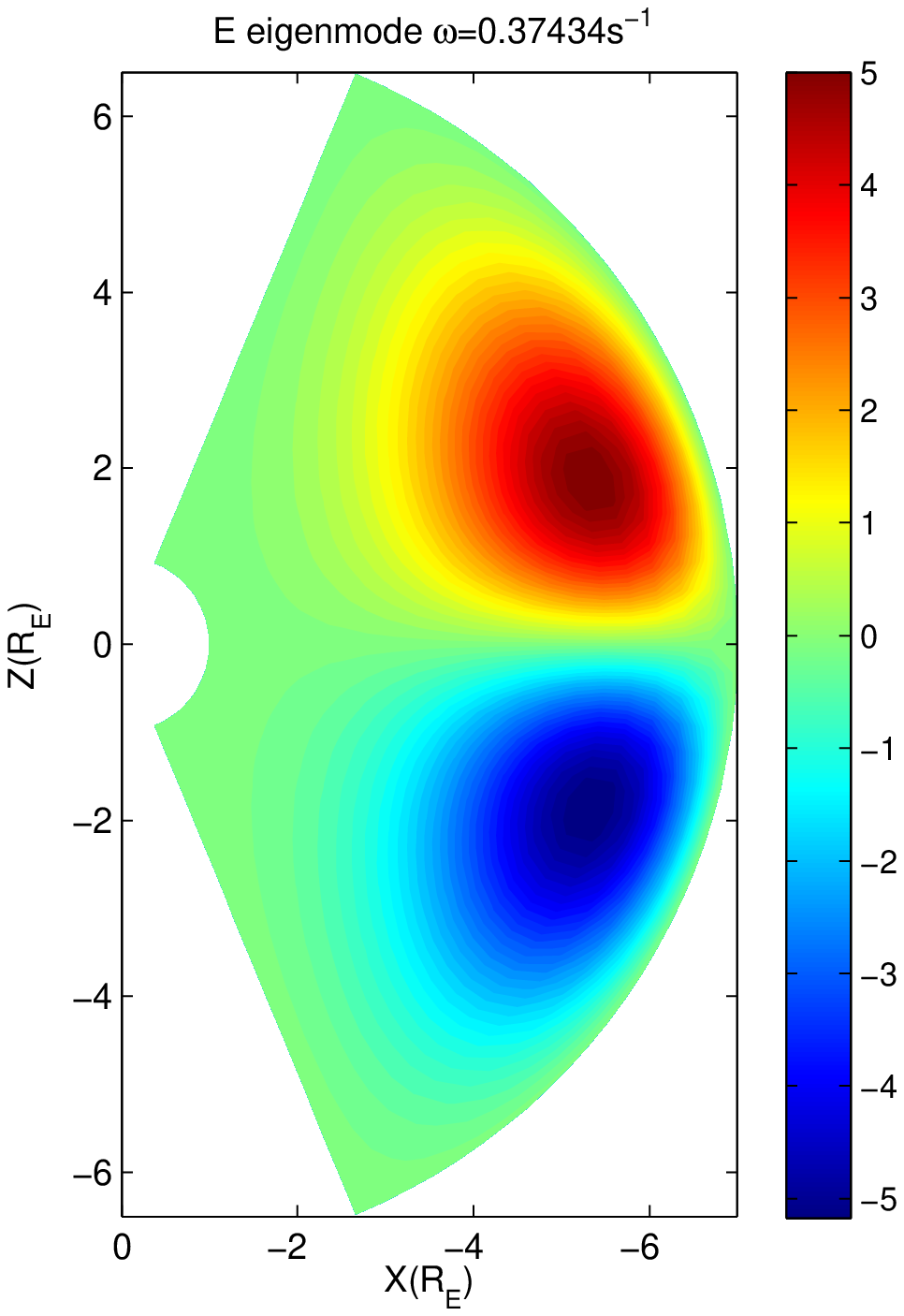}
\includegraphics[width=.329\textwidth]{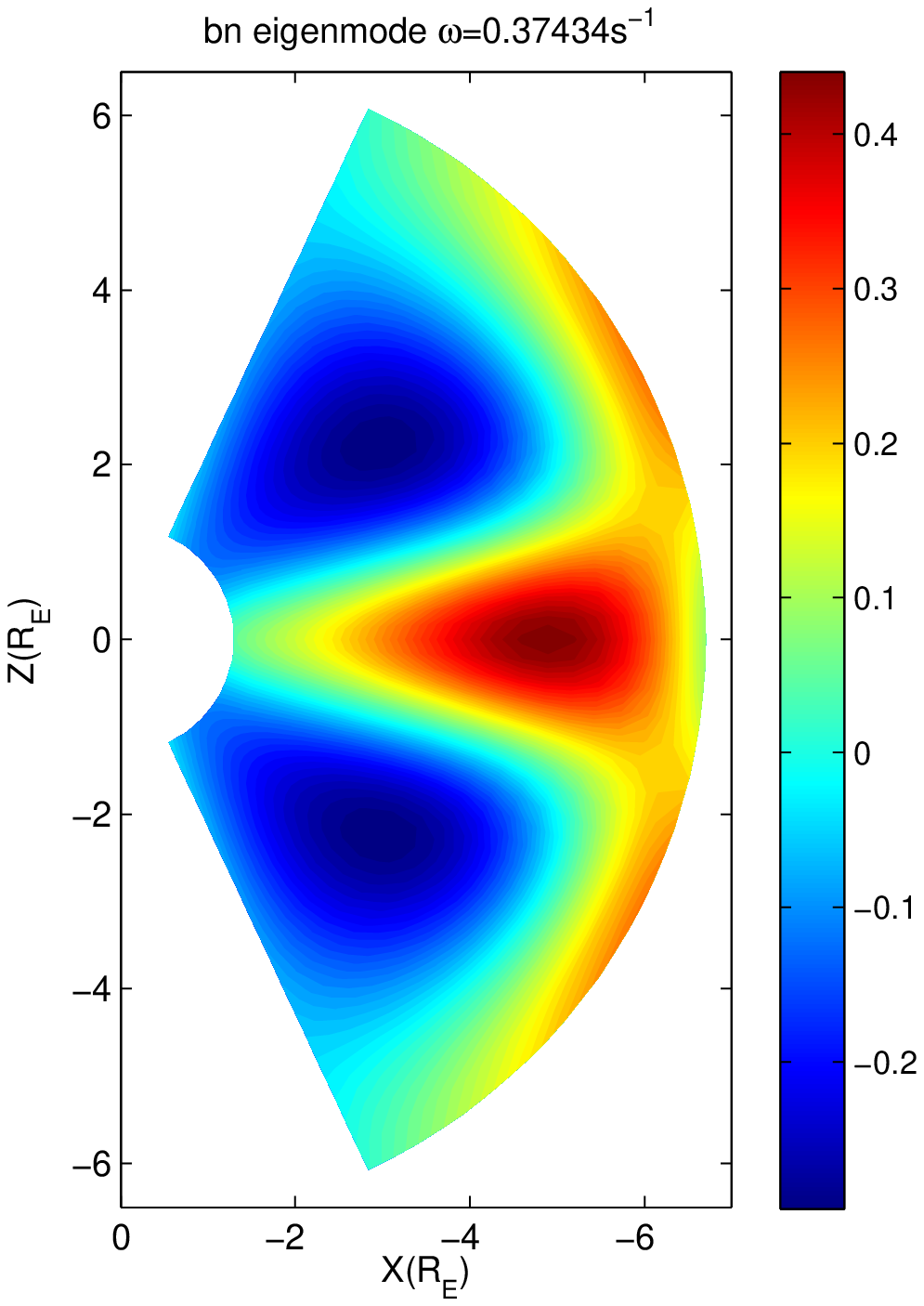}
\includegraphics[width=.329\textwidth]{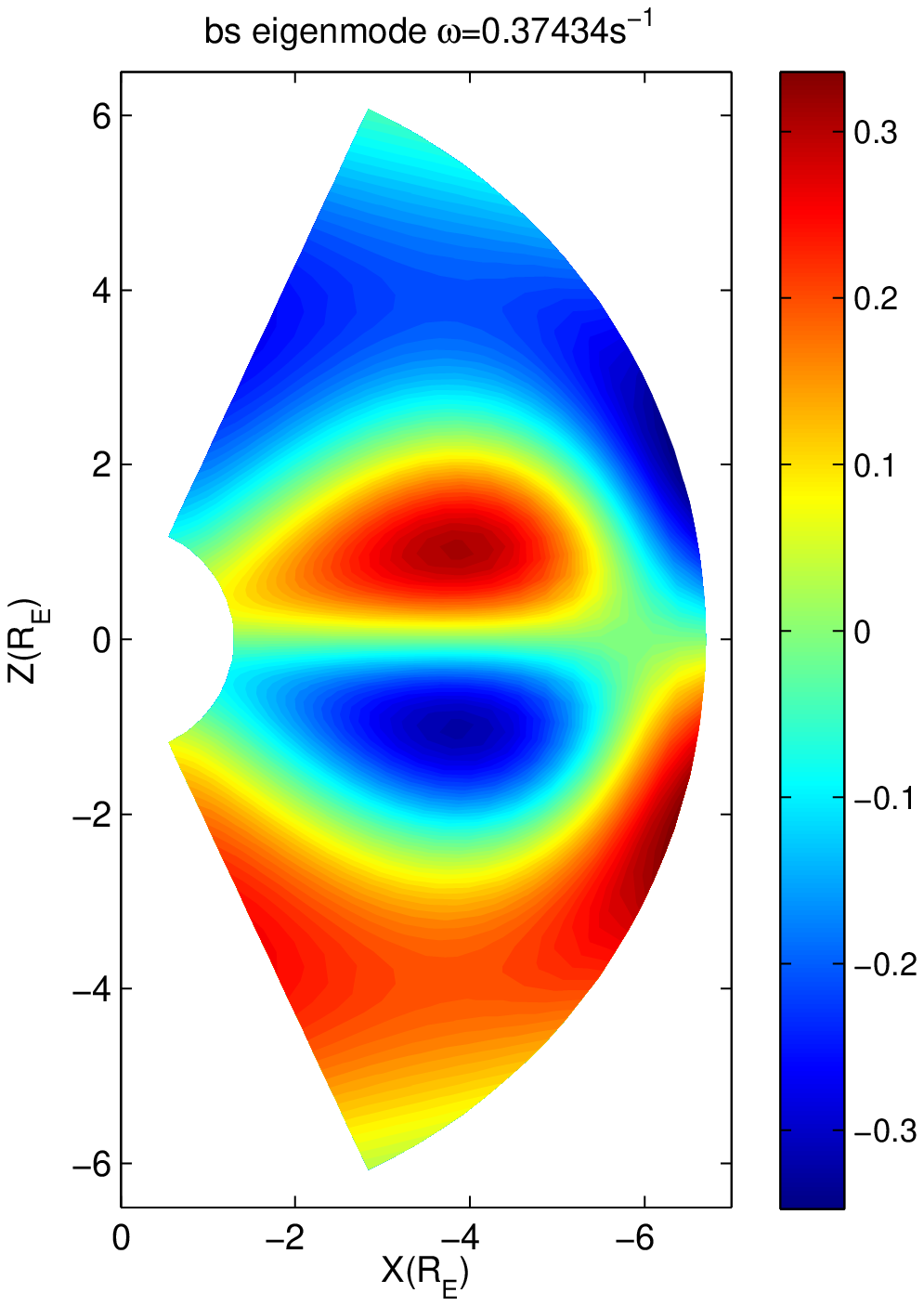}\\
\end{center}
\caption{The eigen-mode solutions of the dipole field. Format is
the same as Fig.~\ref{fignoon}. }\label{figdipole}
\end{figure*}

In Figs.~\ref{fignoon}-\ref{figdipole}, we show the corresponding
eigen-mode solutions of the two lowest-value eigen-frequencies for
the noon, midnight side of the compressed dipole field, and the
standard dipole field, respectively. The solutions generally
display a regular pattern of nodal structures, especially in
higher-order mode solutions of progressively larger
eigen-frequencies (not shown). Starting from the lowest-order
mode, the number of nodes between adjacent extreme values
increases by 1, especially in the $\theta$ direction and sometimes
in the $r$ direction as well. The ``peaks" and ``valleys" are
alternating in appearances as deep red and deep blue patches. The
number of nodes in the normal magnetic field perturbation $b_n$ is
usually increased by one compared with the corresponding electric
field perturbation $E_\phi$. However the number of nodes in the
tangential magnetic field perturbation $b_s$ remains the same as
that of $E_\phi$, albeit the overall shape of the distribution is
much different. Therefore in the fundamental mode solutions, the
normal magnetic field perturbation (representative of  the radial
component) exhibits a two-lobe structure in amplitude symmetric
about the equator, especially in the mid-range  radial distances,
excluding the boundary effect. The tangential component (parallel
to the background magnetic field), on the other hand, shows a
single-lobe structure peaking at the equator. These results are
qualitatively similar to those shown by \citet{2006GMSL}, where
much more sophisticated numerical approaches, and more realistic
background density distribution and boundary conditions were
utilized. Our preliminary numerical experiments also indicate that
the background magnetic field greatly affects the eigenvalue
solutions. For example, the overall behavior of the solutions of
the noon-side compressed dipole is very similar to that of the
standard dipole since their magnetic field topologies are close to
each other. However for the midnight-side compressed dipole, the
locations of the extreme values are generally further stretched to
larger radial distances anti-sunward and the shapes are narrower.

\section{Conclusions and Discussion}
In conclusion, we have examined, in a fairly comprehensive manner,
the decoupled toroidal and poloidal mode hydromagnetic waves in
cold plasmas (ULF waves) with applications to the Earth's inner
magnetosphere, represented by a compressed dipole field model in
addition to the standard dipole field. Under certain assumptions,
the decoupled wave equations are recast into the Klein-Gordon (KG)
form along individual magnetic field lines, especially for both
the toroidal and poloidal transverse Alfv\'{e}n waves. We obtain
the spatial profiles of the characteristic parameters in the KG
formulations including the cut-off frequency $\omega_c$ and the
amplitude factor $f$. The former determines the property of the
solution, i.e., whether it is a propagation type where the
eigen-frequency $\omega>\omega_c$ usually occurring near the
equator or an evanescent type where $\omega<\omega_c$ toward the
footpoints. The latter modulates the amplitude of the wave forms
spatially. We obtain the sets of eigenvalue solutions of
increasing eigen-frequencies and number of nodes in the wave forms
for different background magnetic field geometries. The
corresponding wave periods are in the order of $\sim 1$ second to
$\sim 100$ seconds and compare well quantitatively with prior
studies and observations. In particular, we present a case study
of a fundamental poloidal Alfv\'{e}n wave via our approach and
compare our results with a direct spacecraft observation by
\citet{2013GeoRLD}. The wave period ($\sim 60$s) and the amplitude
ratio ($\sim$ 0.21 nT/(mV/m)) obtained from our approach agree
well with the spacecraft measurements (84s and 0.3 nT/(mV/m),
respectively) and the results from the other approach. For
completeness, we also carry out preliminary numerical computations
of the eigen-mode solutions to the compressional poloidal waves
where the tangential magnetic field perturbation does not vanish.
Qualitatively consistent results with prior studies are obtained.

We acknowledge that the present investigations reported here have
largely been studied in many prior works. The main intellectual
merits thus lie in the aspect of the unique approach via the KG
equations for both the toroidal and poloidal transverse Alfv\'{e}n
waves for a given background magnetic field geometry. The
implications of our results, for example, that of the cut-off
frequency, need to be further explored. In particular, the case
study of a direct comparison with spacecraft measurements yields
promising results, despite the relatively simple and idealized
assumptions about the axi-symmetric geometry and boundary
conditions. It is worth persuing beyond the limitations of the
present approach, especially in conjunction with the
state-of-the-art spacecraft observations, such as those returned
from the Van Allen Probe. Furthermore, it is also desirable to
extend the applications to the solar coronal loop oscillations as
we did in \citet{2012AAH}.




\section*{Acknowledgements}
JFMcK acknowledges support from the Pei-Ling Chan Chair of Physics
in the University of Alabama in Huntsville and the NRF of South
Africa. The other authors acknowledge partial support of NSF grant
AGS-1062050 and NASA grant NNX12AH50G. JZ especially acknowledges
partial support of a graduate research assistantship by the NASA
grant. We are also grateful to Prof. G. Sewell  for his help with
the PDE2D software package.

 \bibliographystyle{elsarticle-harv}
\bibliography{refKG}

\newpage


%
%
%
\begin{table}[htb]
\caption{List of parameters for $L=2,4,6$ of a standard dipole
field.}\label{tbl:dpl} \vskip4mm \centering
\begin{tabular}{ccccc}
\hline
$L$ & $\omega_0$, s$^{-1}$ & $T_0$, s & $\theta_0$, $^\circ$ & $r_0$, $R_E$ \\
 \hline
2 & 0.95 & 6.6 & 49 & 1.15 \\
4 & 0.20 & 32 & 52 & 2.48 \\
6 & 0.085 & 74 & 53 & 3.81 \\
 \hline
\end{tabular}
\end{table}

\begin{table}[htb]
\caption{List of parameters for $L=2,4,6$ of a compressed dipole
field for $\phi=0$ (noon).}\label{tbl:cdpl0} \vskip4mm \centering
\begin{tabular}{ccccc}
\hline
$L$ & $\omega_0$, s$^{-1}$ & $T_0$, s & $\theta_0$, $^\circ$ & $r_0$, $R_E$ \\
 \hline
2 & 0.96 & 6.6 & 50 & 1.20 \\
4 & 0.22 & 29 & 58 & 3.06 \\
6 & 0.10 & 60 & 57 & 5.05 \\
 \hline
\end{tabular}
\end{table}

\begin{table}[htb]
\caption{List of parameters for $L=2,4,6$ of a compressed dipole
field for $\phi=\pi$ (midnight).}\label{tbl:cdpl1} \vskip4mm
\centering
\begin{tabular}{ccccc}
\hline
$L$ & $\omega_0$, s$^{-1}$ & $T_0$, s & $\theta_0$, $^\circ$ & $r_0$, $R_E$ \\
 \hline
2 & 0.94 & 6.7 & 48 & 1.11 \\
4 & 0.18 & 34 & 45 & 1.86 \\
6 & 0.060 & 105 & 83 (39) & 5.81 (1.90) \\
 \hline
\end{tabular}
\end{table}

\begin{table}[htb]
\caption{List of parameters for $L=2,4,6$ of a dipole field for
poloidal standing Alfv\'{e}n waves.}\label{tbl:pol} \vskip4mm
\centering
\begin{tabular}{ccccc}
\hline
$L$ & $\omega_0$, s$^{-1}$ & $T_0$, s & $\theta_0$, $^\circ$ & $r_0$, $R_E$ \\
 \hline
2 & 0.76 & 8.2 & 57 & 1.41 \\
4 & 0.15 & 42 & 62 & 3.09 \\
6 & 0.062 & 101 & 62 & 4.71 \\
 \hline
\end{tabular}
\end{table}

\appendix
\section{KG Formulations for a Compressed Dipole Field}
In this section, we provide the necessary formulations of the KG
equations for a compressed dipole field in a spherical coordinates
with the magnetic field components given by Eqs.~(\ref{eqBr}) to
(\ref{eqBph}).

Since the field remains planar (i.e., $B_\phi=0$), it is possible
to derive the field line equation in each meridional plane ($r$,
$\theta$, $\phi\equiv Const$) with $r$ normalized by Earth radius:
\begin{equation}
\frac{dr}{d\theta}=\frac{rB_r}{B_\theta}=\frac{r(2/r^3-D)}{1/r^3+D}\cot\theta,
\end{equation}
where $D\equiv \frac{b_1}{B_0}(1+b_2\cos\phi)$. This leads to
\begin{equation}
H\sin^2\theta=\frac{r}{D\cdot r^3-2},
\end{equation} where $H$ is an integration constant. As usual, if
we define $r=L$ when $\theta=\pi/2$ ($L$-shell), we obtain
$H=L/(D\cdot L^3-2)$. Then the co-latitude $\theta_F$ of  one of
the footpoints of one particular $L$-shell is obtained by
$\sin^2\theta_F=1/(H\cdot D-2H)$.

Subsequently, the necessary coefficients and factor in the KG
equation for the velocity perturbation in the toroidal mode are
obtained as follows
\begin{eqnarray}
\frac{1}{l_u}&=&-\frac{1}{l_b}=\frac{3}{1+D\cdot r^3}\cot\theta \\
\frac{1}{L_u}&=&\frac{7-4D\cdot r^3-2D^2r^6}{(1+D\cdot
r^3)^2}\cot\theta.
\end{eqnarray}
Additional coefficients such as $1/M_u$ can be derived according
to \citet{2012JGRAW} which enables the derivation of an analytic
form of the cut-off frequency $\omega_c$ appearing in the KG
equation and the amplitude factor $f(\theta)$, relating the
solution of KG equation to the original physical perturbation
quantity.

For the poloidal mode, the  analytic formulae for the relevant
factors in the corresponding KG equation have to be evaluated
numerically since an explicit form of the function $g(\theta)$
from Eq.~(\ref{eq:gpol}) along an individual field line is
difficult to obtain.

\end{document}